\documentclass[showpacs,twocolumn,amssymb,pre,aps]{revtex4}
\usepackage{graphicx,epsfig}
\usepackage{color}

\begin{document}
\title{Thermotropic Orientational Order of Discotic \textbf{Liquid Crystals} in Nanochannels:\\ An Optical Polarimetry Study and a Landau-de Gennes Analysis}

\author{Andriy~V.~Kityk$^{1,2}$ }
\author{Mark~Busch$^{1}$}
\author{Daniel~Rau$^3$}
\author{Sylwia~Ca{\l}us$^{2}$ }
\author{Carole V. Cerclier$^{4}$}
\author{Ronan Lefort$^{4}$}
\author{Denis Morineau$^{4}$}
\author{Eric Grelet$^{5}$}
\author{Christina Krause$^6$}
\author{Andreas Sch\"onhals$^6$}
\author{Bernhard Frick$^7$}
\author{Patrick~Huber$^{1,3}$}
\email[E-mail: ]{andriy.kityk@univie.ac.at, patrick.huber@tuhh.de}

\affiliation{$^1$ Materials Physics and Technology, Hamburg University of Technology (TUHH), D-21073 Hamburg, Germany\\
$^2$ Faculty of Electrical Engineering, Czestochowa University of Technology, 42-200 Czestochowa, Poland\\
$^3$ Experimental Physics, Saarland University, D-66041 Saarbruecken, Germany\\
$^4$ Institut de Physique de Rennes, CNRS UMR 6251, Universit\'e de Rennes 1, 35042 Rennes, France\\
$^5$ Centre de Recherche Paul-Pascal, CNRS UPR 8641, Universit\'e de Bordeaux 1, 33600 Pessac, France\\
$^6$ BAM Federal Institute for Materials Research and Testing, D-12205 Berlin, Germany\\
$^7$ Institut Laue-Langevin, 6 Rue Jules Horowitz, F-38000 Grenoble, France\\}

\date{\today}

\begin{abstract}
Optical polarimetry measurements of the orientational order of a discotic liquid crystal based on a pyrene derivative confined in parallel-aligned nanochannels of monolithic, mesoporous alumina, silica, and silicon as a function of temperature, channel radius (3 - 22 nm) and surface chemistry reveal a competition of radial and axial columnar order. The evolution of the orientational order parameter of the confined systems is continuous, in contrast to the discontinuous transition in the bulk. For channel radii larger than 10~nm we suggest several, alternative defect structures, which are compatible both with the optical experiments on the collective molecular orientation presented here and with a translational, radial columnar order reported in previous diffraction studies. For smaller channel radii our observations can semi-quantitatively be described by a Landau-de Gennes model with a nematic shell of radially ordered columns (affected by elastic splay deformations) that coexists with an orientationally disordered, isotropic core. For these structures, the cylindrical phase boundaries are predicted to move from the channel walls to the channel centres upon cooling, and vice-versa upon heating, in accord with the pronounced cooling/heating hystereses observed and the scaling behavior of the transition temperatures with channel diameter. The absence of experimental hints of a paranematic state is consistent with a biquadratic coupling of the splay deformations to the order parameter.
\end{abstract}


\maketitle

\section{Introduction}

\begin{figure}[tbp]
\epsfig{file=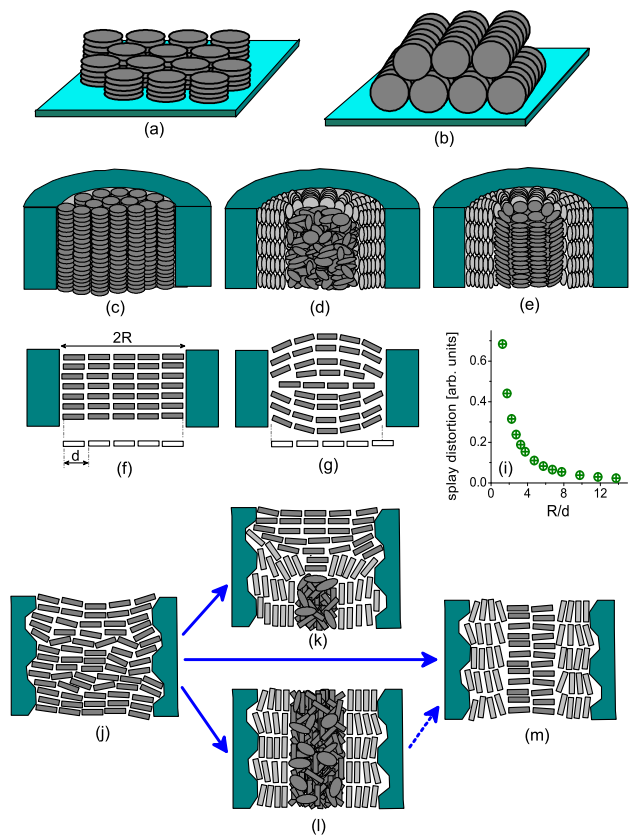, angle=0,
width=0.95\columnwidth}\caption{(color online). Molecular ordering types of discotic liquid crystals. (a) Homeotropic (face-on) ordering on a free surface. (b) Homogeneous planar ordering on a free surface. (c) Parallel axial configuration inside a channel. (d) Radial axial configuration inside a channel with an isotropic core. (e) Escaped radial configuration (combination of radial axial cofiguration near the channel wall and a parallel axial configuration in the core region). (f) undeformed parallel axial configuration inside the channel with a smooth wall. (g) laterally deformed parallel axial configuration inside the channel with a smooth wall. (i) splay deformation vs normalized channel diameter. (j) laterally deformed parallel axial configuration inside the channel with a rough wall. (k) coexsistence of parallel axial and radial axial configuration or (l) radial configuration with isotropic core or (m) escaped radial configuration.} \label{fig1}
\end{figure}

Molecular assemblies consisting of disc-like molecules with an aromatic core and aliphatic side chains exhibit a particularly rich phase transition behavior \cite{Laschat2007, Sergeyev2007, Bisoyi2010}. Because of the $\pi-\pi$ overlap of their aromatic cores they can stack into columns, which in turn arrange in a two-dimensional crystalline lattice leading to thermotropic discotic columnar crystals (DLCs). Thermal fluctuations give rise to liquid-like properties \cite{Haverkate2011} and even glassy disorder can occur \cite{Krause2012, Schoenhals2014}. Therefore, DLCs are particularly interesting systems in order to address fundamental questions in soft matter science, such as structure-dynamics and structure-phase transition relationships.

\begin{figure}[tbp]
\epsfig{file=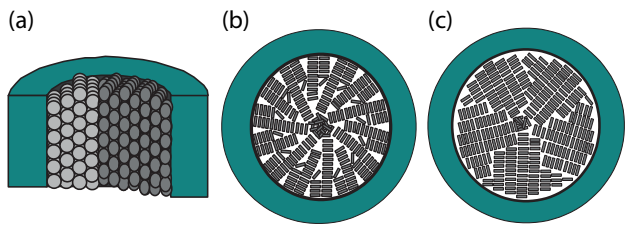, angle=0,
width=0.95\columnwidth}\caption{(color online). Schematics of radial molecular ordering of discotics with long-range, hexagonal columnar order confined in cylindrical channels. (a) Sideview on two columnar domains. (b) Topview on a radial structure with dislocation defects and a disordered core. (c) Sideview on a radial multi-domain state with disordered core.} \label{fig:DefectStructure}
\end{figure}

DLCs encompass also advantageous materials properties, including highly anisotropic visible light absorption, long-range self-assembly, self-healing mechanisms, high charge-carrier mobilities along the column axis and a tuneable alignment of the columns \cite{Grelet2006, Laschat2007, Sergeyev2007, Feng2009}. Therefore they represent promising systems for active layers in organic devices, such as field-effect transistors and photovoltaic cells.

The combination of DLCs with mesoporous solid matrices offers the opportunity to prepare nanowires and to design hybrid systems \cite{Steinhart2005, Duran2012} with interesting opto- and opto-electronic properties. However, similarly as the thermotropic behaviour of rod-like liquid crystalline systems \cite{Crawford1996, Binder2008, Kityk2008} the properties of confined discotic systems has turned out to be particularly susceptible to nano confinement and to be substantially altered in comparison to the bulk systems \cite{Kopitzke2000, Steinhart2005, Stillings2008}. 

For discotics the number of available experimental and theoretical studies under confinement is much smaller than for rod-like liquid crystals. Therefore, the goal of this paper is the exploration of the thermotropic orientational arrangement of a discotic liquid crystal in a selected set of mesoporous matrices (alumina, silica, silicon) as a function of channel diameter and depending on the chemical treatment of the channel surfaces (as prepared, silanized, covered by ink) by high resolution optical polarimetry. These experimental investigations are discussed with respect to a phenomenological Landau-De Gennes model in order to arrive at mechanistic insights in the thermotropic phase behavior of the system.


\section{Experimental}

The discotic molecule used in our study was the pyrene 1,3,6,8-tetracarboxylic rac-2-ethylhexyl ester (Py4CEH). It consists of a polyaromatic core surrounded by flexible aliphatic chains. Synthesis details of this molecule are described in Ref. \cite{Hassheider2001}.

{ The porous alumina membranes ($p$Al$_2$O$_3$) were purchased from Smart Membranes GmbH (Halle, Germany). { The channel radii and porosities have been determined by volumetric N$_2$-sorption isotherms at $T$ = 77~K:} $R = 21.0 \pm$2 nm (porosity $P$=24\%, thickness $h=$100~ $\mu$m), 15.5$\pm$1.5 nm ($P=$17\%, $h=$100 $\mu$m), 10.0$\pm$0.7 nm ($P=$16\%, $h=$90 $\mu$m), 7.8$\pm$0.5 nm ($P=$16\%, $h=$80 $\mu$m). 

To obtain membranes of porous silica ($p$SiO$_2$), electrochemically etched free standing silicon membranes ($p$Si) \cite{Gruener2008} were subjected to thermal oxidation for 12 hours at $T=$800 $^o$C under standard atmosphere. By a variation of the etching time we obtained $p$SiO$_2$ membranes with average channel radius 3.4$\pm$0.2 nm ($P$ = 13\%, $h=$100~$\mu$m), 3.8$\pm$0.2 nm ($P$ = 15\%, $h=$120~$\mu$m ), 4.6$\pm$0.3~nm ($P$ = 30\%, $h=$180~$\mu$m ), 5.7$\pm$0.4~nm ($P$ = 35\%, $h=$280~$\mu$m ), 6.8$\pm$0.5~nm ($P$ = 46\%, $h=$310~$\mu$m ),  and a $p$Si membrane with channel radius 8.5$\pm$0.5~nm ($P$ = 65\%, $h=$300~$\mu$m ), as determined by recording of volumetric N$_2$-sorption isotherms at $T$ = 77~K. The mesoporous membranes were completely filled by capillary action (spontaneous imbibition) of LCs in the isotropic phase \cite{Huber2007, Gruener2009, Gruener2011}. For the bulk measurements, the sample cells, made of parallel glass plates ($h$ = 11~$\mu$m ), were filled with the discotic LC. The temperature scans were performed with a heating and cooling rate of 0.12~K/min}

The optical polarization setup employs a photoelastic modulator and a dual lock-in detection system, as described in Ref. \cite{Kityk2009}. It provides an accuracy of the optical retardation measurements that is better than 5$\cdot$10$^{-3}$~deg. The sample normal, which coincides with the long channel axis, was tilted out by an angle of 40-45$^o$ with respect to the laser beam. Most measurements have been performed by means of a He-Ne laser with a wavelength $\lambda$=633 nm, where almost no optical absorption exists. An exception are the investigations of the $p$Si membranes ($h=$300 $\mu$m), which are not suitable for optical polarization measurements at this wavelength due to strong light absorption. In this case, the measurements were performed in the infrared region, at $\lambda$ = 1342 nm for which $p$Si is optically well transparent. The refractive anisotropy in liquid crystal materials and the related optical birefringence result from collective molecular rearrangements of positional and/or orientational type. In principle, DLCs exhibit both of these ordering types. However, the contribution to the optical birefringence because of positional ordering, e.g. because of a collective hexagonal translational order, is much smaller than the orientational one. To simplify the analysis we will therefore ignore the positional rearrangement in the following. In uniaxial discotic liquid crystals the degree of orientational order can be described by the scalar order parameter $S=\frac{1}{2}\langle 3\cos^2\phi-1\rangle$, where $\phi$ is the angle between the axis perpendicular to the polyaromatic plane of a discotic molecule and a direction of preferred local molecular orientation (director). The brackets denote an averaging over all molecules under consideration. Optical polarimetry is a particularly suitable technique to explore orientational order, since it scales linearly with $S$.

\section{Results and discussion}

We recall that native surfaces of silica glass, likewise the inner surfaces of alumina, silica or silicon membranes exhibit hydrophilic properties enforcing face-on type ordering of Py4CEH, see Fig.~\ref{fig1} a) \cite{Brunet2011}. Accordingly a cell consisting of parallel glass plates ($h$=11 $\mu$m) filled with this system leads to perfect homeotropic alignment. In a confined cylindrical geometry this anchoring type may lead to a radial columnar (or radial nematic) alignment close to the channel walls as it is sketched in Fig.~\ref{fig1}d or Fig.~\ref{fig1}e. Leaving aside the ordering in the core region of the channels, which shall be discussed later, we emphasize that columnar type ordering was indeed confirmed in recent small-angle neutron scattering and X-ray diffraction studies on Py4CEH embedded in anodized alumina membranes with channels in a diameter range of $D$=25 to 50 nm \cite{Cerclier2012}. Note, however, that the variation of the integrated Bragg intensities typical of the columnar order exhibit a systematic decrease with channel diameter in that study. A simple extrapolation of this trend predicts a vanishing of the columnar structure for channel sizes smaller than 10 nm, which was experimentally proved by the absence of columnar order in porous silicon with a channel diameter of 8 nm.

This observation can be traced to the topological frustration arising from the incompatibility between the homeotropic anchoring condition at the channel wall and the cylindrical channel symmetry. Independently of the channel size the curvature of the molecular lamella increases while approaching the channel axis. This can result in splay distortions or other defect structures (see Fig. \ref{fig:DefectStructure}) which hamper or even inhibit the growth of ordered columnar domains, see \ref{fig1}d. Such defects are also expected to lower the local transition temperature as we discuss in a more quantitative manner in the second part of the manuscript for the case of splay distortions. In particular, the topological frustration may simply bring the core region into a disordered (isotropic) state (Fig.~\ref{fig1} d) or other energetically more favorable structures, like e.g. escaped radial configurations as sketched in Fig.~\ref{fig1}e. In fact, the escaped radial configuration has been experimentally observed for rod-like nematic liquid crystals confined into cylindrically shaped sub-micron channels imposing similarly competing geometric constraints \cite{Crawford1991, Crawford1996}.


\begin{figure*}[tbp]
\epsfig{file=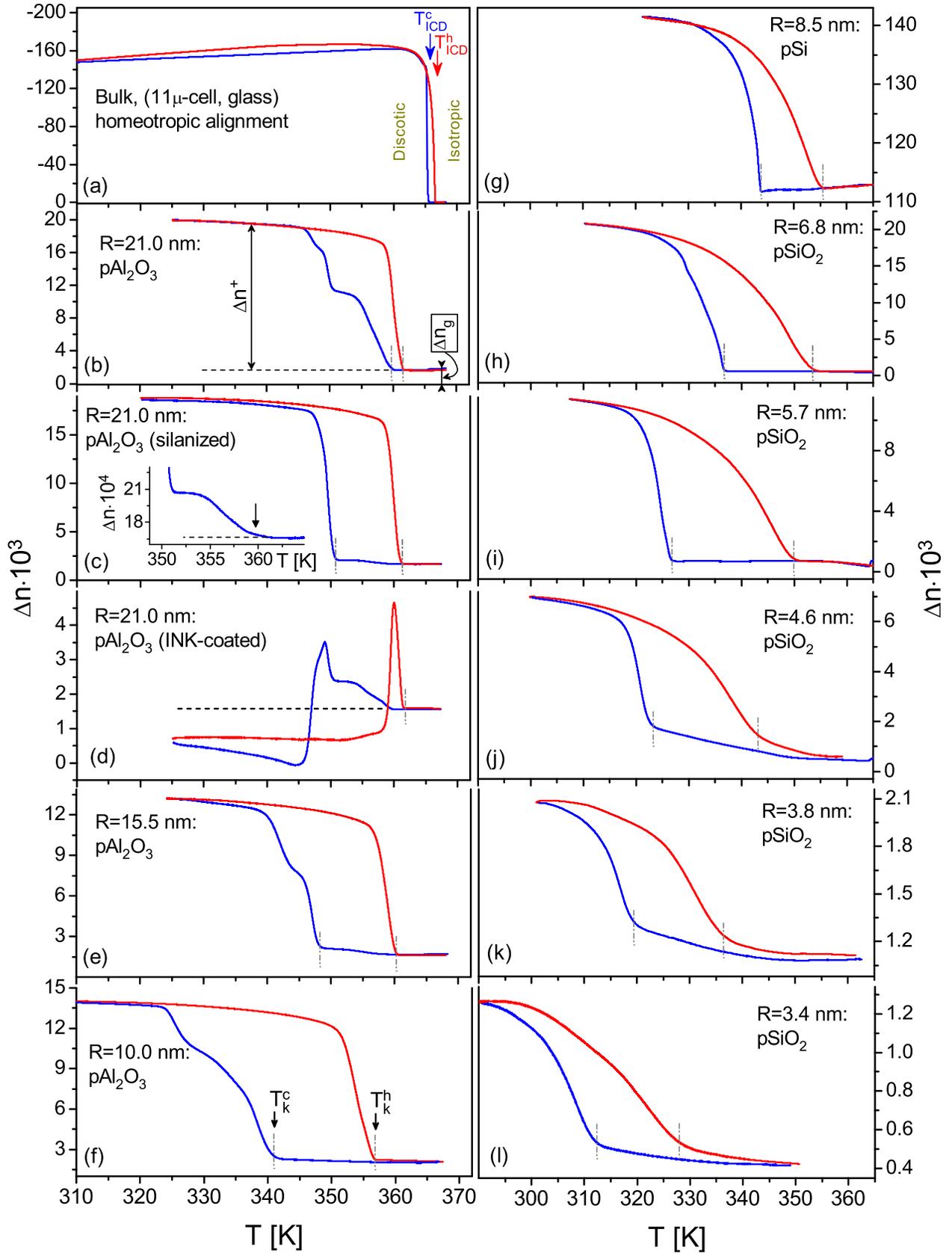, angle=0,width=1.95\columnwidth}
\caption{(color online).  The optical birefringence vs. temperature during cooling (blue online color) and subsequent heating (red online color). (a) bulk Py4CEH, (b)-(l) - the birefringence of confined Py4CEH embedded in different nano channels as indicated in the figure. The quantity $\Delta n_g$ and $\Delta n^+$ labeled in panel (b) mark the geometrical and excess birefringence, respectively. The arrows in panel (a), labeled $T_{IDC}^c$ and $T_{IDC}^h$, mark the bulk phase transition temperatures from the isotropic-to-columnar discotic phase and vice-versa, respectively. The gray vertical dash-dot lines, labeled $T_k^h$ and $T_k^c$, mark the characteristic temperature of kink position during heating and cooling runs, respectively.} \label{fig2}
\end{figure*}


\subsection{Results}

In Fig.~\ref{fig2} we present the temperature dependences of the optical birefringence $\Delta n$ measured in heating (red online color) and cooling (blue online color) runs for discotic Py4CEH in the bulk state and embedded in mesoporous matrices of different type, channel size, and chemical treatment.


\subsubsection{bulk}
In the homeotropically aligned, columnar phase bulk Py4CEH exhibits \emph{negative} birefringence. The columnar-to-isotropic transition is accompanied by a jump-like vanishing of $\Delta n$, see Fig. \ref{fig2}a. Subsequent cooling reveals a small temperature hysteresis ($\sim$1 K) of the reverse transformation. The temperature variation of the optical birefringence in the columnar discotic phase exhibits saturation. In fact, during cooling to ambient temperature the absolute magnitude of the bulk birefringence even decreases slightly. Probably, this unusual behavior can be attributed to the changes in the positional ordering. The decrease of the interdisc and intercolumnar distances upon cooling\cite{Cerclier2012} provide a weak positive contribution on top of the considerably larger negative birefringence resulting from the saturated homeotropic orientational ordering. Note that the value of approx. $\Delta n=$-0.15 is in good agreement with the birefringence value for $\lambda=633$ nm reported in Ref. \cite{Charlet2008}, where wavelength dependent measurements on
Py4CEH-films with unidirectional planar alignment are reported.

\subsubsection{Spatial confinement in nanochannels}
For the confined Py4CEH the orientational order inside the channels results in an excess birefringence, $\Delta n^+$, which appears on the background of the geometric birefringence, $\Delta n_{\rm g}$ typical of the parallel-aligned nanochannels, see the arrow in Fig.~\ref{fig2}b and the Eqs. (1)-(3) of Ref. \cite{Kityk2009}. The measured birefringence represents, therefore, a superposition of these two contributions, i.e., $\Delta n = \Delta n_{\rm g} + \Delta n^{\rm+}$.  The geometric birefringence depends on the difference between the refractive indices of the host matrix and the guest material. Therefore it is relatively small for entirely filled silica or alumina membranes, however quite large for the silicon membrane. This can be inferred from the socket values in Fig.~\ref{fig2}. While the geometric birefringence $\Delta n_{\rm g}$ exhibits smooth and comparably weak temperature variations, caused by varying refractive indices of host and guest materials as well as changes of the porosity due to thermal expansion, the excess birefringence $\Delta n^+$ is characteristic of the thermotropic molecular orientational ordering and dominates the measurements as can be seen below.

For membranes with untreated surfaces the excess birefringence is \emph{positive}, see Fig.~\ref{fig2}. This indicates a dominating radial face-on type molecular ordering. Thus, it is in good agreement with the aforementioned diffraction studies \cite{Cerclier2012}. In contrast to bulk Py4CEH, which exhibits a first-order isotropic-to-columnar discotic phase transition the geometrical constraint renders the transition in the nanochannels gradual with characteristic kinks at $T=T_{\rm k}^{\rm c}$ (cooling) and $T=T_{\rm k}^{\rm h}$ (heating), see vertical gray dash-dot lines in each panel. {  Please note that it will become evident from the discussion below that the characteristic temperatures  $T=T_{\rm k}^{\rm c}$ and $T=T_{\rm k}^{\rm h}$ do not represent phase transition temperatures in a classical meaning, as e.g. in the case of a bulk state. They rather characterize a local transition temperature in the molecular layer(s) close to the channel wall.} Above  $T_{\rm k}^{\rm h}$ ($T_{\rm k}^{\rm c}$) the birefringence changes weakly with temperature. Just below these characteristic temperatures it exhibits strong variations with a saturation at lower temperatures. The larger the channel diameter, the steeper are the $\Delta n$-decays and -increases. Spatial confinement also strongly influences the temperature hysteresis. It substantially increases as the channel diameter decreases. Moreover, for large channel radii ($R \ge 10.0$ nm) the birefringence measured at cooling exhibits below $T_{\rm k}^{\rm c}$ a staircase-like variation. By contrast, upon heating the optical birefringence gradually decreases starting from a saturated value down to the pure geometrical birefringence, $\Delta n_g$ for all systems investigated.

The axial parallel molecular ordering (see Fig.~\ref{fig1}a), which appears to be desirable for a number of organic electronic applications, never occurred in our experiments as long as we dealt with untreated membranes. This fact is evident, since the native hydrophilic surfaces of alumina and silica enforce a homeotropic (face-on) type ordering. Porous silicon is also no exception in that respect. It is characterized by a native layer of SiO$_2$ and its anchoring properties, as has been experimentally proved recently \cite{Calus2012}, are the same as for the silica membranes. Silanization, on the other hand, renders these surfaces more hydrophobic enforcing thus an unidirectional planar ordering, see Fig.~\ref{fig1}b. The experiment, however, shows that even silanized channel walls are not able to stabilize the axial parallel molecular order, although a certain effect of silanization can be observed in the channels of large diameters, compare e.g. Fig.~\ref{fig2}b and Fig.~\ref{fig2}c. We emphasize that parallel axial and radial types of ordering, as it is sketched in Fig.~\ref{fig1}c and Fig.~\ref{fig1}d, respectively, provide an excess birefringence of opposite signs. Thus these configurations are easily distinguishable by our optical polarimetry technique. We find here that silanization increases the temperature hysteresis of the isotropic-to-columnar discotic transition, only. A more detailed inspection of this region, see inset in Fig.~\ref{fig2}c, reveals two kinks at cooling. A weak positive excess birefringence, which appears just below 360 K, could be attributable to a nonuniform molecular ordering with coexisting regions of radial and parallel axial configurations, where a radially ordered component slightly prevails. Below 351 K, however, the birefringence exhibits a strong positive increase indicating a formation of a strongly dominated radial configuration.

Coating of the inner surface with gelatine based ink appears to be more efficient in a stabilization of axial ordering, see Fig.~\ref{fig1}b.  But even in this case the regions of parallel axial configuration coexist or compete with regions of a radial molecular configuration, see in Fig.~\ref{fig2}d. At cooling the excess birefringence increases first slightly, but then decreases to a value lower than the level of the geometrical birefringence $\Delta n_g$ (see extrapolated dashed line in Fig.~\ref{fig2}d). This means that the negative contribution caused by the parallel axial ordering slightly dominates in this case over the positive contribution due to the radial arrangement of discotic molecules. The coexisting state is stable at least down to 320 K. At subsequent heating the coexisting configurations are stable up to about 358 K.  Above this temperature, at about 360 K, first the regions with parallel axial configuration disappear, as is evident from a strongly birefringence rise, and at even slightly higher temperatures, at about 361.5 K,  the regions of radial configuration vanish.

\subsection{Discussion}
In the following we extract characteristic quantities of our polarimetry measurements in order to allow for a more quantitative evaluation and discussion of our experiments.

\subsubsection{Phase transition temperature and birefringence versus channel diameter}
\begin{figure}[tbp]
\begin{flushleft}
\epsfig{file=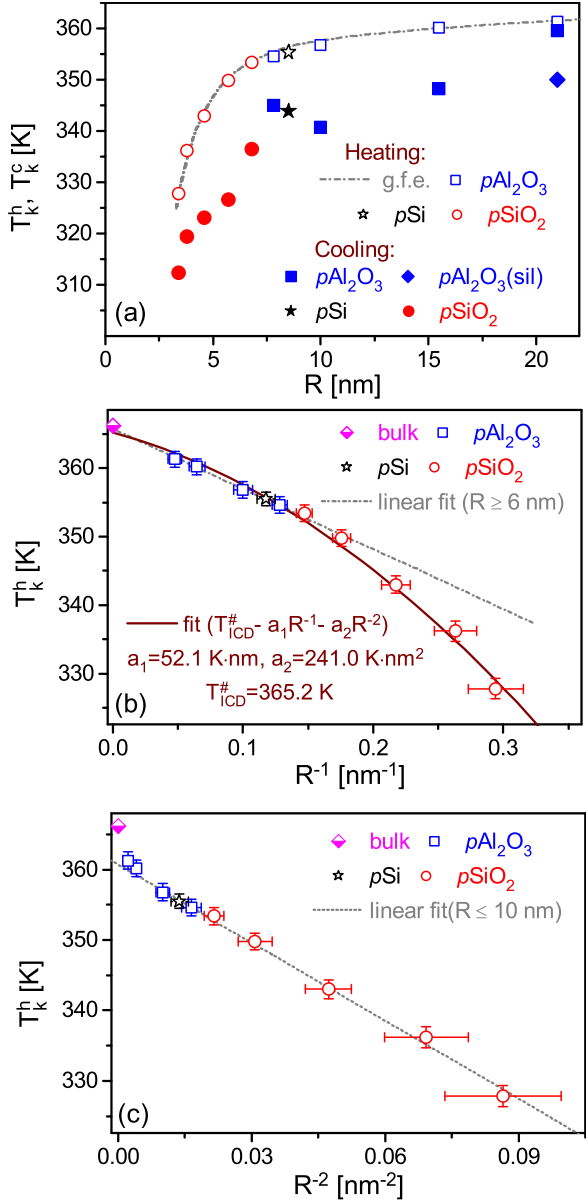, angle=0, width=0.9\columnwidth}
\caption{(color online). Panel (a): The temperatures of the kink positions, $T_{\rm k}^{\rm h}$ and $T_{\rm k}^{\rm c}$, determined from the temperature dependences of the optical birefringence (see Fig.~\ref{fig2}) measured during heating and cooling, respectively, versus the channel radius $R$. Panel (b): $T_k^h$ versus the inverse channel radius, $R^{-1}$. Open symbols with error bars are the experimental data points measured at heating. The dashed-dot line (gray online color) is the best linear fit for the channel radii larger than 6 nm. Solid curve (wine online color) represents the best fit by the function $T_{\rm k}^{\rm h}=T_{\rm ICD}^{\#}-a_1R^{-1}-a_2R^{-2}$ with the fit-parameters $T_{\rm ICD}^{\#}=$ 365.2 K, $a_1=$ 52.1 K$\cdot$nm and $a_2=$ 241.0 K$\cdot$nm$^2$. Panel (c): Kink position $T_{\rm k}^{\rm h}$ versus the inverse square of the channel radius, $R^{-2}$. The broken line is the best linear fit for the data points that correspond to channel radii $R \le 10$ nm.} \label{fig3}
\end{flushleft}
\end{figure}

In Fig.~\ref{fig3}a the $R$-dependence of the temperature positions of the kinks, $T_{\rm k}^{\rm h}$ and $T_{\rm k}^{\rm c}$ are plotted. Both temperatures vary nonlinearly with the channel radius, $R$. However, $T_k^h(R)$ changes smoothly in contrast to $T_{\rm k}^{\rm c(R)}$. This indicates that the structural transformations upon heating and cooling follow considerably different pathways. In particular it is interesting to note that $T_{\rm k}^{\rm h}$ depends on the channel radius only, i.e. it is the same for membranes with native and coated surfaces of the same channel size. This is in strong contrast to the observation in cooling runs, where the kink position and the overall evolution of structural ordering strongly depends on the surface anchoring.

\begin{figure}[tbp]
\begin{flushleft}
\epsfig{file=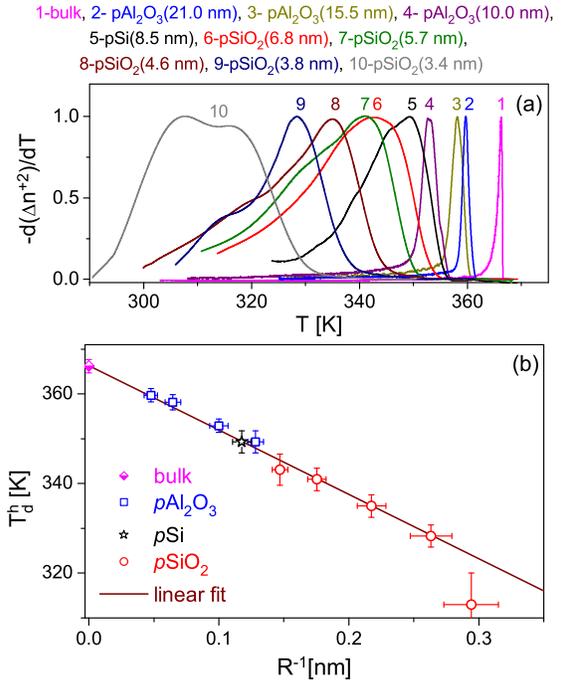, angle=0, width=0.85\columnwidth}
\caption{(color online). { Panel (a): The normalized temperature derivatives of the excess birefringence squared, $d(\Delta n^{+2})/dT$, versus $T$ determined from the temperature dependences of the optical birefringence (see Fig.~\ref{fig2}) measured during heating; All the derivatives are normalized to their maximum value. Panel (b): The temperature position of the derivative maximum versus the inverse channel radius, $R^{-1}$. Solid line (wine online color) is the linear fit. The data point corresponding to $R=$3.4 nm is excluded from the fitting procedure.}} \label{fig3-2}
\end{flushleft}
\end{figure}

{ An interesting and challenging question arises with respect to the $R$-scaling of the characteristic transition points, particularly the kink temperature position, $T_{\rm k}^{\rm h}(R)$.
In the case of melting or freezing in pores the shift of the transition temperatures with respect to the bulk transition is expected to scale according to the Gibbs-Thomson equation with $1/R$ \cite{Christenson2001, Wallacher2001, Alba-Simionesco2006, Knorr2008}. In panels (b) and (c) of Fig.~\ref{fig3} we show $T_{\rm k}^{\rm h}$ vs $R^{-1}$ and $R^{-2}$, respectively. When the entire range of radii is considered, $T_{\rm k}^{\rm h}$ versus $R^{-1}$ evidently exhibits a nonlinear dependence. For large channel radii, $R \ge 6.8$ nm, however, the experimental data fit the linear dependence, see gray dash-dot line in Fig.~\ref{fig3}b and an extrapolation of this dependence to $R^{-1}=0$ gives the bulk transition temperature, $T_{\rm ICD}^{\rm h}$.  The $T_{\rm k}^{\rm h}(R^{-2})$-dependence, on the other hand, can be well fitted by a linear dependence in practically the entire range of channel radii, see gray broken line in Fig.~\ref{fig3}c. However, an extrapolation of such a linear fit to $R^{-2}=0$ gives a  temperature of about 5.5 K lower than the bulk temperature $T_{\rm ICD}^{\rm h}$. Hence, neither the $R^{-1}-$ nor the $R^{-2}-$ scaling can solely describe the $R$-shift of the kink temperature $T_{\rm k}^{\rm h}(R)$: We observe a crossover from a $R^{-1}$-dependence at larger channel radii to a $R^{-2}$-scaling for small channel radii. In the entire range of channel radii the $T_{\rm k}^{\rm h}(R)$-dependence can be described by $T_{\rm k}^{\rm h}=T_{\rm ICD}^{\#}-a_1R^{-1}-a_2R^{-2}$ with the fit-parameters $T_{\rm ICD}^{\#}=$ 355.2 K, $a_1=$ 52.1 K$\cdot$nm and $a_2=$ 241.0 K$\cdot$nm$^2$. The parameters$a_1$ and $a_2$ are material specific, since they depend both on the intermolecular forces of the discotic molecules and on the interaction of the discotic molecules with the porous host material.  


One must emphasize again that the characteristic point $T_{\rm k}^{\rm h}$ corresponds according to our understanding to the local transition that takes place in the molecular layer(s) next to the channel wall. Therefore, it differs from the characteristic transition temperature as determined by differential scanning calorimetry (DSC), which corresponds to the fastest temperature variations of the effective (averaged over the pore volume) order parameter squared, i.e. the maximum of the derivative $d(\langle S \rangle^2)/dT$ \cite{Kityk2010}. Hence, considering that the excess birefringence $\Delta n^+ \propto \langle S \rangle$, the temperature position of the DSC anomaly should coincide with the temperature position of the characteristic anomalous peak in $d(\Delta n^{+2})/dT$ \cite{Kityk2010}. Fig.~\ref{fig3-2}a presents
$-d(\Delta n^{+2})/dT$, normalized to their maximum, versus $T$ as determined from the temperature dependences of the optical birefringence (see Fig.~\ref{fig2}) measured during heating. For large channel diameters their temperature dependence is characterized by a relatively sharp peak being located a little bit below the kink position. For smaller channel radii the peak substantially broadens and shifts considerably down with respect to the kink position. Interestingly the temperature behavior of the derivative $-d(\Delta n^{+2})/dT$ for the membrane with smallest channel radius, $R=3.4$ nm, is characterized by a very broad anomaly with overlapping double peaks. This may be caused by a bimodal channel diameter distribution, nonuniform channel diameter and/or channel wall roughness. Therefore, we ignore this measurements in the further analysis. The temperature position of the derivative maximum, $T_{\rm d}^{\rm h}$, measured at heating versus the inverse channel radius, $R^{-1}$, is shown in Fig.~\ref{fig3-2}b. It indicates a $R^{-1}$-scaling in agreement with recent DSC measurements on this system \cite{schonhals2013}. Albeit, here we prove the $R^{-1}$-scaling for the effective (averaged) phase transition point to a considerably smaller channel radii, i.e. for $R\ge3.8$ nm.}

\begin{figure}[tbp]
\begin{flushleft}
  \epsfig{file=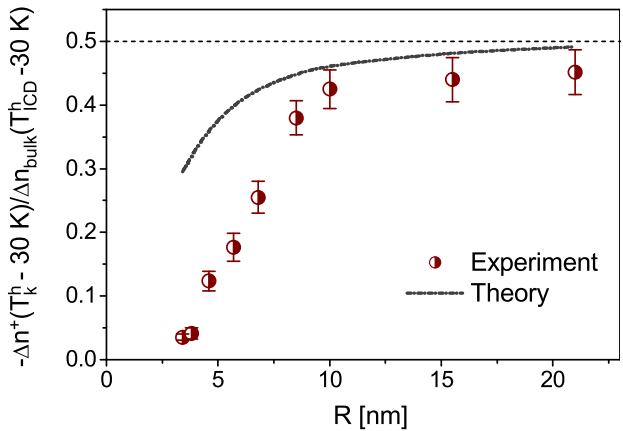, angle=0, width=0.95\columnwidth}
  \caption{(color online). The ratio between the saturated excess birefringence, $\Delta n^+$, of the confined discotic liquid crystal Py4CEH, taken at $T_{\rm k}^{\rm h}-30K$, and the saturated bulk birefringence at $T_{\rm ICD}^{\rm h}-30K$ as function of channel radius, $R$. Semiopen circles (wine online color) are the magnitudes obtained in the experiment. Dash-dot curve (dark gray online color) are the calculated values for a radial configuration with isotropic core. The horizontal, dashed black line is the  theoretical limit.} \label{fig4}
\end{flushleft}
\end{figure}

The excess birefringence $\Delta n^+$ scales with the porosity of the membrane, $P$. Accordingly, a comparison of the orientational ordering in samples of different porosity with the bulk state requires its normalization by $P$. To characterize a saturated ordering it is useful to compare the normalized excess birefringence inside the channel, $\Delta n^{+*}=\Delta n^+/P$, and the saturated bulk birefringence, $\Delta n^*_{\rm bulk}$, as the reference value. For the fully developed (saturated) parallel axial ordering the normalized
excess birefringence is approximately equal to the saturated bulk birefringence,
i.e. $\Delta n^{+*}(\rm axial)=\Delta n^*_{\rm bulk}$. For the fully developed radial
ordering, on the other hand, the normalized excess birefringence equals
half of the saturated normalized axial birefringence and it is opposite in
sign, see the discussion on page 4, right column in Ref. \cite{Chahine2010}, i.e. the following relation holds $\Delta n^{+*}(\rm radial)=-\Delta n ^{+*}(\rm axial)/2 = -\Delta n^{*}_{\rm bulk}/2$.
At $T \le T_{\rm k}^{\rm h}-30 K$ the excess birefringence, $\Delta n^+$  in all matrices investigated saturates, i.e. $\Delta n^+(T_{\rm k}^{\rm h}-30$K$)\approx \Delta n^{+*}$. It is worthwhile to compare this quantity with the bulk birefringence value, $\Delta n_{\rm bulk}$ taken at $T_{\rm ICD}^{\rm h}-30$ K in order to gain information on the extent of molecular order at low temperature. In Fig.~\ref{fig4} the ratio $-\Delta n^+(T_{\rm k}^{\rm h}-30$K$)/\Delta n_{\rm bulk}(T_{\rm ICD}^{\rm h}-30$K$)$ is plotted versus the channel radius. For large channel diameters ($2R\ge20$ nm) this ratio is $\sim$ 0.45, i.e. it is only about 10\% smaller than 0.5, corresponding to a fully developed radial ordering. However, for channels with smaller diameters this ratio is considerably smaller. It monotonically decreases with decreasing channel diameter, down to 0.035 for a silica membrane with $R=$3.4 nm. Obviously the radial ordering inside the channels is not uniform.

On the one hand, following our recent structural study \cite{Cerclier2012} the molecular ordering near the channel walls is known to be either of radial columnar (for large channel diameters, Bragg peak typical of hexagonal order observable) or radial nematic type (for small channel diameters, Bragg peak typical of hexagonal order not observable). In both cases the radial orientational order provides positive contribution to the excess birefringence. On the other hand, we do not know much about the translational and orientational molecular ordering close to the channel axis, i.e. in the core region. Our experiments are compatible with two possibilities: The discotic molecules in the core region can be either orientationally disordered, giving no contribution to the excess birefringence, or they are aligned preferably parallel to the channel axis, providing thus a negative contribution to the excess birefringence. Those two alternatives are sketched in Fig.~\ref{fig5}a and Fig.~\ref{fig5}b, respectively. Presumably, the interface between the shell, consisting of radially ordered discotic molecules, and the core is smooth. One probable configuration is shown in Fig.~\ref{fig5}b, where the director changes gradually its spatial orientation from radial (at the channel wall) to parallel axial (in the channel center), which corresponds to an escaped radial configuration.

\subsection{Phenomenology of the Isotropic-to-columnar discotic transition in cylindrical channels}
In the following a series of phenomenological and geometrical considerations will be presented in order to arrive at an appropriate interpretation of the observations presented above. They are based on the extensive knowledge existing for defect structures of liquid crystals and our previous diffraction studies on this discotic system in nanochannels. 

Our measurements indicate that in none of the investigated cases a well developed parallel axial configuration can be achieved, even in the case when surface anchoring enforces this type of ordering in the channels. Why is this the case? Some hints for answering this question may be found in pure geometrical reasons. It is imaginable that the the hexagonal lattice accommodates to the cylindrical constraints by having undulations of the columns (with some bend elastic energy)\cite{Oswald2006}. In principal, a fully developed parallel axial configuration requires, however, that the channel diameter is commensurate with the lateral intercolumnar distance, i.e. an integer number of columns should fit exactly into a circular section of the channel, as sketched in Fig.~\ref{fig1}c or Fig.~\ref{fig1}f. Otherwise, a splay distortion results (see Fig.~\ref{fig1}g) which rises the free energy and possibly renders an alternative radial configuration with isotropic core (Fig.~\ref{fig1}d) or escaped parallel configuration (Fig.~\ref{fig1}e) more favorable. Assuming an ideal circular cross-section, the splay distortion equals zero if $2R=n \cdot d$. It reaches a maximum value for $2R=(n-1/2) d$, here $n$ is an integer number and $d$ is the equilibrium intercolumnar distance, see Fig.~\ref{fig1}f. For the condition, $2R=(n-1/2)d$, simple geometrical considerations yield the equation, $(n-1/2)u_sd/2=\sin(nu_sd/2)$. Solving this equation numerically for different $n$ one obtains a decaying dependence for the splay deformation, $u_s(R/d)$, see Fig.~\ref{fig1}i.

The intercolumnar distance of Py4CEH as determined from the position of the leading, hexagonal (100)-Bragg peak is $d =2/\sqrt{3} \cdot 2 \pi / q_{\rm 100}=$2.04 nm \cite{Cerclier2012}. { Accordingly, approximately 3 ($2R=6.8$ nm) up to about 21 ($2R=42.0$ nm) columns may fit into the radial cross-section of the channels investigated here.} From Fig.~\ref{fig1}i it follows that more chances for a realization of that configuration have the membranes with large channel diameters, since the splay distortion significantly decreases with the channel radius $R$. This agrees with our experimental observation.

The channel wall roughness additionally reduces chances for the realization of the pure parallel axial configuration (see illustration in Fig.~\ref{fig1}j). Particularly, for channels with small mean diameters the corresponding distortions are spatially inhomogeneous (due to the roughness induced channel diameter variations) and strong, rendering the parallel axial configuration unstable and giving way for the radial configuration with isotropic core (Fig.~\ref{fig1}l) or escaped radial configuration (Fig.~\ref{fig1}m).

For larger channel diameters, on the other hand, the splay distortions caused by the geometrical restraint are considerably smaller thus at adequate anchoring conditions this may lead either to a fully developed parallel axial structure (never observed in our experiments) or coexisting regions with the parallel axial and radial configurations, as e.g. sketched in Fig.\ref{fig1}k. The latter is presumably the case for the surfactant ink-coated membranes, at least a "mixed" type of ordering is consistent with the negative excess birefringence observed in this experiment.

As mentioned already above for channel radii larger than 10 nm a Bragg peak typical of the translational, hexagonal arrangement of the LCs in confinement could be observed, whereas for smaller pore diameters this signature was not detectable. Therefore we suggest that for larger pore diameters either uniaxially, aligned columns (see Fig. 3) or radial, aligned columns with dislocation defects are established. Note that splay deformations are not compatible with the positional columnar order observed for the larger channel diameter, only with the nematic-like organization for the small pore diameter.

\subsection{Uniaxially aligned columns or radial columns with dislocations defects}

In the first case the DLC is uniaxially aligned in the direction normal to the nanochannel axes. This mainly radial configuration without elastic distortion preserves the hexagonal lattice but exhibits unfavored alignment of the discotic molecules at the DLC-solid interface and possibly the coexistence of domains (see Fig. \ref{fig:DefectStructure}). The anchoring energy is minimized for the face-on (i.e. homeotropic) orientation with respect to the side-on (planar) orientation, even if the difference of energy has been  reported to be relatively small \cite{Grelet2006}, $\gamma^{\parallel}_{\textrm{\scriptsize{DLC-solid}}}-\gamma^{\bot}_{\textrm{\scriptsize{DLC-solid}}}\approx10^{-5}~J/m^{2}$. The different channels in a given membrane being independent, all the radial orientations are averaged in space and are therefore relatively consistent with the scattering experiments showing mainly intercolumnar positional order along the nanochannel direction \cite{Cerclier2012}. Moreover, one can imagine that along the long axis of the channels the azimuthal orientation of domains of aligned columns changes, also in agreement with the scattering results. 

In the second case an increasing number of dislocations (and/or grain boundaries) (see Fig. \ref{fig:DefectStructure} ) is necessary in order to fulfill simultaneously the geometrical constraints of the nano channels and the anchoring conditions. Unfortunately, we do not have more detailed experimental data in order to distinguish between those two, possible structures. 

For the pores with smaller diameters we assume a radial discotic state with possible short-range, but with no long-range hexagonal order, i.e. a nematic type of order, which adapts by splay deformations to the confined geometry.

\subsection{Nematic, radial discotic state}
\subsubsection{Splay deformations and Landau-De Gennes model }
In the following we aim at a more quantitative description of the confined radial discotic state without translational order. To this end we refer to a phenomenological Landau-de Gennes analysis of nematic discotic order subjected to elastic distortions as outlined in Ref.\cite{Skacej1998} and explore to what extent our observations can be qualitatively and quantitatively captured by such a model. The free energy density of a nematic state in the case of elastic distortions can be expressed as:

\begin{eqnarray}
f&=&f_0+f_c,  \nonumber \\ \label{eq1}
f_0&=&\frac{1}{2}A(T-T^*)S^2-\frac{1}{3}BS^3+\frac{1}{4}CS^4, \\
f_c&=&\frac{1}{2}\{b_1(\vec{\nabla}\cdot\vec{n})^2+b_2[\vec{n}\cdot(\vec{\nabla}\times\vec{n})]^2+b_3[\vec{n}\times(\vec{\nabla}\times\vec{n})]^2 \quad \nonumber \\
&-&b_4\vec{\nabla}[\vec{n}(\vec{\nabla}\cdot\vec{n})+\vec{n}\times(\vec{\nabla}\times\vec{n})]\}S^2 \nonumber
\end{eqnarray}

where $A$, $B$, and $C$ are the Landau-de Gennes free energy expansion coefficients, $T^*$ is an effective temperature. The quantity $f_0$ represents the free energy density of the unperturbed liquid crystal in the ground state and $f_c$ is the free energy density describing the coupling between the nematic order parameter $S$ and different types of elastic distortions with coupling constants $b_1$-$b_4$.

\begin{figure}[tbp]
\epsfig{file=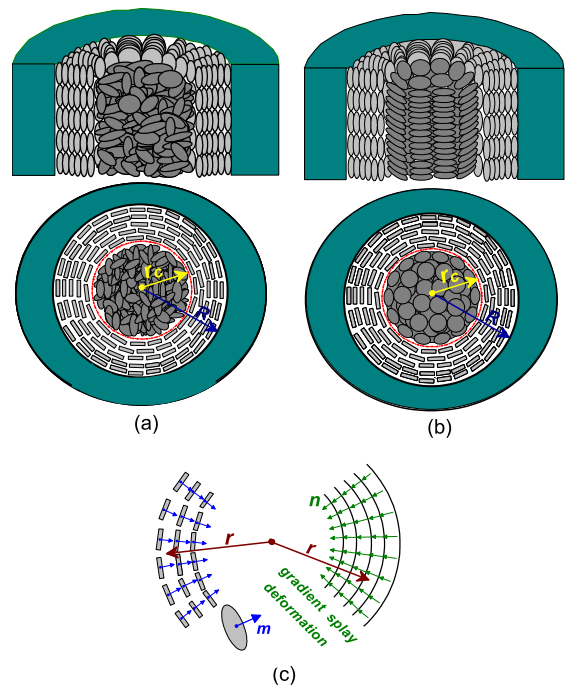, angle=0, width=0.85\columnwidth}
\caption{(color online). Radial configuration with isotropic core (a) and escaped radial configuration (b). Panel (c): The left sectorial fragment represents the molecular ordering, the right sectorial fragment explores the splay distortion due to a curvature of molecular layers with the local director, $\vec{n}$, oriented radially. The resulting splay distortion, $\vec{\nabla}\vec{n}$, equals $1/r$, i.e. it is spatially inhomogeneous. The gradient in the splay distortion results in a phase front (see red circle of radius $r_c$), which separates a radially ordered shell from an isotropic core. According to our model calculation it moves during cooling from the periphery to the channel center, and vice-versa during heating.} \label{fig5}
\end{figure}

In order to employ this model for our confined system, we have to analyze which type of elastic distortions are present for different assumed configurations of orientational order in the channels. While the radial columnar (or radial nematic) ordering near the channel walls (i.e. in the shell region) has been well established in neutron experiments \cite{Cerclier2012}, a challenging question remains the structural configuration in the core region. In general, it may be disordered or parallel axial ordered, as it is sketched in Fig.~\ref{fig5}a and Fig.~\ref{fig5}b, respectively. Although the neutron scattering experiments have not revealed any hints of an existence of an escaped radial configuration with columnar parallel axial ordering in the core region (Fig.~\ref{fig5}b), one cannot exclude an existence of such a phase.

Nevertheless, we resort here to the most simple case of a radially ordered shell with an isotropic core, i.e. the configuration shown in Fig.~\ref{fig5}a and do not consider the possibility of more complicated orientational order distributions. Moreover, we ignore the spatial inhomogeneities of the scalar order parameter ($\vec{\nabla}S$-terms) considered in Ref. \cite{Skacej1998} for the sake of simplicity.  Hence we assume that its spatial variation in consequent molecular cylindrical layers of the nematic shell is weak. This is justified by a strongly saturated temperature behavior of the scalar order parameter $S(T)$ observed in the bulk state, which presumably originates in the strong $\pi-\pi$ interaction between the polyaromatic cores. Consequently the gradient terms are omitted in the free energy density (see Equation \ref{eq1}.

For the radially ordered configuration with isotropic core the gradients of the order parameter occur most likely in the interface regions, e.g. near the channel walls and/or on the isotropic-nematic interface (phase front). However, we believe that those effects do not significantly influence the molecular ordering in neighboring molecular layers, again because of the saturation of the nematic molecular ordering. In other words, in our simplified model with isotropic core, see Fig.~\ref{fig5}a, the radial dependence of the nematic order parameter $S(r)$ is approximated as, $S(r)$=0 at $r < r_c$ and $S(r)\approx 1$ at $r_c < r \le R$.


For a cylindrical geometry with a radial arrangement of discotic molecules in a nematic shell, as it is sketched in Fig. \ref{fig5}a, the bend ($\vec{n}\times(\vec{\nabla}\times\vec{n})$) and twist ($\vec{n}\cdot(\vec{\nabla}\times\vec{n})$) distortions both equal zero, whereas the splay deformation, $\vec{\nabla}\cdot\vec{n}=1/r$, i.e. is inversely proportional to the distance from the channel axis $|\vec{r}|$. { The free energy density reads then:

\begin{equation}
\label{eq2}
f(R,r)=\frac{1}{2}A(T-T^*(R)+b_1A^{-1}/r^2)S^2-\frac{1}{3}BS^3+\frac{1}{4}CS^4,
\end{equation}
i.e. it is $r$-dependent. By solving $f=0$ and $\partial f/\partial S=0$, one obtains a local temperature of the isotropic-to-nematic transition, which  has a local  ($r$-dependent)  character. It thus renders the transition gradual in accordance with the experimental
observation:
\begin{equation}
\label{eq3}
T_c(r)=T^*(R)+2B^2/(9AC)-b_1A^{-1}/r^2,
\end{equation}
Here we consider the competition of volume free energies (scaling with $1/R^2$) and interfacial free energies (scaling with $1/R$) which results to the depression of the effective transition temperature, $T^*(R)$ as $R^{-1}$ (Gibbs-Thomson mechanism with cylindrical phase boundary). Accordingly, it can be approximated  as $T^*(R)=T^*(R\rightarrow \infty)-g/R$ where $g$ represents a material constant. The effective temperature $T^*(R)$ depends on the channel diameter. We neglect, however, the channel diameter distribution in our sample, thus is is assumed to be constant in the entire pore volume. By contrast the third term of the equation (\ref{eq3}), $b_1A^{-1}/r^2$, depends on $r$, which results in a radial dependence of the transition temperature. From Eq. \ref{eq3} it follows that this ''local'' transition temperature remains constant within the cylindric layer of radius $r \le R$, but varies in subsequent cylindrical molecular layers from the channel center ($r\rightarrow 0$) to the periphery ($r\rightarrow R$). At $r,R\rightarrow\infty$ it approaches the bulk transition temperature, $T_{ICD}=T^*+2B^2/(9AC)$.} For cylindrical layers close to the channel axis, i.e. $r\rightarrow 0$, the local transition temperature $T_c(r)$ is considerably shifted down. Thus our model calculation yields an important prediction with regard to the spatial evolution of the phase transition: The phase transition starts near the channel walls upon cooling. Furthermore, it is accompanied by a cylindrical phase front, which gradually propagates from the periphery to the channel center upon cooling, and vice-versa upon heating. The cylindrical front of radius $r_c$ (see Fig.~\ref{fig5}a) separates the radially ordered nematic phase ($r_c<r<R$) from the isotropic core region ($r<r_c$). Note that within such an approach a disordered core region of final  { radius $r_c=[AT^*(R)/b_1+2B^2/(9Cb_1)]^{-1/2}$ }is omnipresent, even for $T \rightarrow 0$, which results from the diverging splay deformation in the channel centre.  The phenomenological model evidently fails for small core radii on the order of the diameter of a single molecule, a value close to the intercolumnar distance of approx. 2 nm \cite{Cerclier2012}.

The analysis above allows for an understanding of our experimental observations. During cooling the transition from the isotropic-to-nematic phase starts at the periphery, near the channel walls. Therefore, the interfacial interaction and the channel roughness play an important role. Radial homeotropic and axial planar ordering may simultaneously occur, enforced by specifics of the interfacial and/or local geometrical conditions. Accordingly first a (metastable) coexistence of both states appear and only at lower temperature the radial molecular ordering becomes dominating. This explains the observation that the kink position upon cooling varies from sample to sample and is sensitive to the surface coating, compare e.g. Fig.~\ref{fig2}b and Fig.~\ref{fig2}c.

By contrast, the kink position during heating exhibits a more systematic evolution as a function of channel diameter.  Presumably, there is always an entirely disordered or highly defect decorated core. It acts as a nucleation for the high-temperature phase. The phase front starts there and gradually propagates according to the $r$-dependent transition temperature (Eq 2) to the periphery, while "melting" subsequent radially ordered molecular layers. The kink position at heating corresponds therefore to the situation at which the phase front reaches the channel wall. { It is defined mainly by a geometrical factor, i.e. the channel radius. For channels of radius $R$ with ideally smooth walls its temperature is given by Eq. \ref{eq3}, i.e. $T_{\rm k}=T_{ICD}-gR^{-1}-b_{\rm 1}A^{-1}R^{\rm -2}$. The material constants here, as derived from the fit-parameters of the kink position measured at heating, are $T_{\rm ICD}=T_{\rm ICD}^{\#}=$ 355.2 K, $g=a_1=$ 52.1 K$\cdot$nm and $b_{\rm 1}A^{-1}=a_2=$ 241.0 K$\cdot$nm$^2$.}

There is another remarkable peculiarity in the phase transition scenario outlined above. Since the size of the isotropic core decreases gradually as the temperature lowers, its diameter can reach an integer number of column diameters at certain temperatures. The splay/bend distortions caused by the geometrical constraint vanish in this case, which could result in a parallel axial ordering in the vicinity of the channel axis. Hence, the director, $\vec{n}$ would change its spatial orientation from a radial, near the channel wall, to a parallel axial orientation near the channel axis (see e.g. Fig.~3b), which corresponds to the escaped radial configuration known from rod-like nematic liquid crystals\cite{Crawford1991, Crawford1996}. Moreover, the molecular configurations may be stable in a certain temperature interval, only. The radially ordered shell tries to induce radial order in the neighbouring core region. This may lead to a new metastable configuration with less number of axially aligned molecules in the core region. One may speculate that the staircase-type ordering phenomenologies observed in a couple of the cooling scans are hints of such mixed radial/axial transition states with well-defined numbers of columns in the channel center.

\subsubsection{Order parameter evolution at the isotropic-to-discotic columnar transition}

Encouraged by the reasonable agreement between our experimental findings and the predictions of the simplified model, we extend now our analysis and calculate the temperature evolution of the optical birefringence, that is the evolution of the nematic order.

We recall that in the case of rod-like systems confined in silica channels a remarkably well agreement with Landau-de-Gennes models for the evolution of nematic order could be achieved in the past \cite{Kutnjak2003, Kutnjak2004, Kityk2008, Kityk2010}. Note, however, that a principle difference exists between the confined axially ordered rod-like nematics considered there and the radially ordered discotics discussed here. For the rod-like systems the confinement effects were mainly represented by an effective ordering field, $\sigma$, which results from the anchoring condition of the molecules at the channel walls. The strength of this field depends on the channel size ($\sigma(r)\propto r^{-1}$), but it is \emph{spatially homogeneous}. Depending on the actual $\sigma$-value, which may be either subcritical ($\sigma<1/2$) or overcritical ($\sigma>1/2$), the transition in a spatially confined geometry is of first or second order, respectively. \emph{Bilinear coupling} between the nematic order parameter $S$ and the ordering field $\sigma$ ($\sigma S$-term) breaks the symmetry of the isotropic state. Therefore there exists a weak residual nematic ordering (called also paranematic state) at temperatures even far above the bulk isotropic-to-nematic phase transition. It is preferentially located in the interface region near the channel walls.
Here we consider a strong gradient in the splay distortion, which radially develops inside the channels. \emph{Biquadratic coupling} between the nematic order parameter $S$ and a splay distortion ($(\vec{\nabla}\cdot\vec{n})^2S^2$-term) obviously does not break the symmetry of the isotropic state. Thus pretransitional effects are not expected. The deformed state of the molecular layers does not change the first order character of the phase transformation. It only extents the temperature range in which the cylindrical front shifts from the periphery to the channel axis or vice-versa.

These considerations are again in good agreement with our experimental results: For channel diameters down to 5.7 nm there are indeed no pretransitional effects. However, for smaller channel radii small tails are observable. They may originate: i) in an increasing influence of spatial inhomogeneity in the channel diameter, what directly leads to a smearing of the kink in the dependence $\Delta n$ vs $T$ and ii) in a rising weight of surface effects resulting from a certain disorder induced in few molecular layers next to a rough channel wall.

\begin{figure}[tbp]
\begin{flushleft}
  \epsfig{file=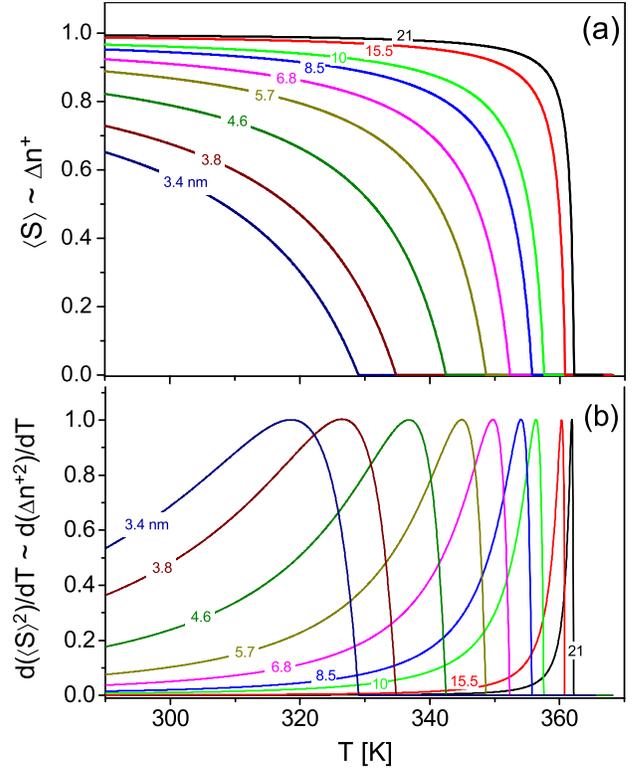, angle=0, width=0.95\columnwidth}
  \caption{(color online). { Panel(a):} The effective order parameter, $\langle S \rangle$ vs $T$, calculated by averaging $S(r)$ over the cylindrical channel volume for the radial ordered configuration with isotropic core (Fig.~7a).  In relevant simulations it is assumed that the temperature dependence of  $S(r)$ exhibits fully saturated spatial variation in consequent cylindrical layers, i.e. $S(r)=0$ at $T \ge T_c(r)$ and $S(r)=1$ at $T < T_c(r)$, where $T_c(r)$ is the local transition temperature. Excess birefringence, $\Delta n^+$, is proportional to $\langle S \rangle$. {  Panel(b): Temperature derivatives, $-d(\langle S \rangle^2)/dT$, normalized to their maximum values versus $T$ as calculated from the $\langle S(T) \rangle$-dependences shown in Panel (a).}} \label{fig6}
\end{flushleft}
\end{figure}

Now, we calculate the temperature behavior of the excess birefringence $\Delta n^+$. It is proportional to the effective order parameter, $\langle S \rangle$, representing the average value of the local order parameter $S(r)$ over the channel volume, i.e. $\langle S \rangle = V^{-1}\int S(r)dV$.  { In our calculations the radial dependence of the local transition temperature is taken as  $T_c(r)=T_{\rm ICD}^{\#} -a_1R^{-1}-a_2r^{-2}$, where $T_{\rm ICD}^{\#}$, $a_1$ and $a_2$ are the fit values as determined above by fitting of the experimental $T_{\rm k}^{\rm h}(R)$-dependence. The dependence $T_c(r)$, defined in this way,  correctly describes the local transition temperature in the molecular layer(s) near the channel wall, it may however somewhat deviate from this value in the higher molecular layers. We believe that this difference is not significant. The core region is assumed to be isotropic, being separated from the radially ordered shell by a sharp boundary. We also assume that the temperature dependence of $S(r)$ exhibits a fully saturated behavior in cylindrical layers, i.e. $S(r)=0$ at $T \ge T_c(r)$ and $S(r)=1$ at $T < T_c(r)$. In such a case the effective order parameter is given by the analytical expression, $\langle S (T)\rangle=1-[r_c(T)/R]^2=1-a_2R^{-2}/(T_{\rm ICD}^{\#}-a_1R^{-1}-T)$.}

In Fig.~\ref{fig6}a the calculated order parameter $\langle S \rangle$ vs $T$ is displayed for a selected set of channel diameters, $R$. It can be directly compared with the temperature variations of the measured optical birefringence shown in Fig.~\ref{fig2}. The simple model reproduces not only the temperature shift of the kink position very well, but also correctly describes the evolution of the curve shape: For larger channel diameters the temperature variations of the effective order parameter are steeper. Thus $\langle S \rangle $ reaches quickly the corresponding saturated level. Moreover, the saturated value of the effective order parameter decreases monotonically with decreasing channel radius. This effect can be attributed to an increasing weight factor of the isotropic core. { Also the calculated temperature derivatives, $-d(\langle S \rangle^2)/dT$ (see Fig.~\ref{fig6}b), exhibit a very asymmetric shape similar to the ones observed in the experiment, see Fig.~\ref{fig3-2}a. The anomalies are quite sharp at large channel diameters, but considerably broaden with decreasing $R$.}

Finally, it is interesting to compare the calculated ratio of the saturated excess birefringence for the discotic nematic in the confined and bulk states, $-\Delta n^+(T_{\rm k}^{\rm h}-30$K$)/\Delta n_{\rm bulk}(T_{\rm ICD}^{\rm h}-30$K$)$ as a function of channel radius $R$ (Fig.~\ref{fig4}, dash-dot line, dark gray online color) with the one derived from the experiment, see Fig. \ref{fig4}. In general the evolution of the experimental data is remarkably well captured. However, in all cases the experiment shows significantly lower values than the calculated one and the deviations considerably increases at small channel diameters. These discrepancies can have several causes: The molecular ordering near the channel walls may be reduced due to interface roughness. Particularly at smaller channel radii the relative contribution of these molecules, not considered in our model, considerably rises. Moreover for smaller channel sizes our continuum model may systematically underestimate the confinement effects. Note that for $R=$3.4 nm only about 4 layers fit into the channel space. Here the simple phenomenological model systematically underestimates the elastic deformations and microscopic molecular dynamics simulations would be the adequate method in order to quantify the confinement effects.


\section{Summary and conclusion}

We presented a systematic optical polarimetry study aimed at an exploration of the thermotropic collective orientational order of discotics confined in parallel-aligned nanochannels of mesoporous alumina, silicon, and silica. In agreement with a former diffraction study the temperature-variation of the optical birefringence indicates collective orientational order, typical of the columnar hexagonal bulk phase. The phase transition temperatures both in cooling and heating decrease with decreasing channel radius, whereas the width of the transition hysteresis increases. The negative magnitude of the birefringence indicates a radial columnar phase for the native and silanized channel surfaces, also in agreement with the previous diffraction experiments. Only for the peculiar case of an ink-coated alumina surface we could achieve parallel axial columnar order. But even there a radial ordered state occurs as a transition state.
Upon cooling we found several hints for a stair-case type ordering, which could be hints of transition states with integer number of axially ordered columns in the centre. Interestingly, this is reminiscent of the quantized evolution of smectic order known to occur at the free surface of rod-like liquid crystals \cite{Ocko1986}.

The existence of long range translational order, as indicated by previous scattering experiments on the identical systems (for $R>10$ nm), and the results here indicate a radial order, which is presumably significantly affected by grain boundaries and dislocation defects in order to allow for translational and orientational order under the confining constraints.

For the smaller pore diameters ($R<10$ nm) where no hints of long-range translational order have been found, we anticipate a radial, nematic order dominated by splay deformations. A Landau-De Gennes model considering elastic splay deformations in cylindrical layers of radially arranged molecule columns and an orientationally disordered core, resulting from competing geometrical constraints in the channel centre, yields semi-quantitative agreement with the observations in our experiments, both with respect to the temperature-dependence of the birefringence and with respect to its evolution as a function of channel diameter. The model predicts a nucleation of the columnar phase at the channel wall and its gradual propagation to the channel centre upon cooling. By contrast upon heating the isotropic state nucleates in the channel centre. The cylindrical phase boundary propagates towards the channel wall. Since the latter process is nucleated in the channel centre, the transition upon melting is more reproducible and predictable, not affected by metastable phases. More generally spoken, the pronounced hysteresis effects observed originate on the different nucleation sites of the low-temperature phase upon cooling and the high-temperature phase upon heating, respectively.

{ The local transition temperature detected in the molecular layer(s) near the channel walls, exhibits two types of scaling: (i) at large channel diameters it scales with $1/R$, in good agreement with the Gibbs-Thomson prediction for a phase transition of first order in confinement; and (ii) for small channel diameters it scales with $1/R^2$, caused by the increasing impact of splay distortions. The effective (averaged) transition temperature, determined from the temperature derivatives of the excess birefringence squared, exhibits $1/R$ scaling in agreement with recent DSC measurements, although our optical studies proves the validity of this scaling up to pore radii $R\ge 3.8$ nm.} The biquadratic coupling of the splay deformations to the order parameter are consistent with the absence of experimental hints of a paranematic state. This is in strong contrast to the nematic order/disorder transition in confined calamatic nematics, where a bilinear coupling with a homogeneous surface field leads to a pronounced paranematic behavior upon nano confinement.

It is also interesting to compare our findings for this disc-like system with melting of spherical building-blocks, e.g. argon in nanochannels. Also for this first-order bulk transition the movement of the ordering/disordering interface has been intensively discussed and experimentally explored in the past \cite{Christenson2001, Alba-Simionesco2006, Knorr2008, Webber2010}. Interfacial melting with a radial moving solid/liquid boundary has been proposed based on filling fraction dependent investigations \cite{Alba-Simionesco2006, Wallacher2001, Schaefer2008, Moerz2012}. Note however, that here the high temperature (liquid) phase is believed to be nucleated at the pore wall and hence the movement of the front boundary is opposite to the scenario outlined here for the melting of the columnar discotic state.


The validity of the Gibbs-Thomson equation is the base of thermoporometry, that is the determination of pore diameter distributions from the temperature shifts of phase transitions \cite{Mitchell2008, Petrov2009, Riikonen2011, Kondrashova2013}. Therefore we want to stress that we provide here an example, where geometrical constraints render the $1/R$-scaling law inappropriate for a conversion of phase transition shifts into pore radii distributions, if one relies purely on the bare optical birefringence measurements. Rather an appropriate conversion of these data sets to a quantity which corresponds to a specific heat signal, that is the temperature-derivative of the square of the birefringence, follows the simple $1/R$-scaling.

Despite the fact that our simple continuum model calculations reasonably well agree with our experimental findings, we believe that molecular dynamic simulations of confined discotics will allow for more detailed comparisons and insights with regard to the confinement effects. In fact such calculation on rod-like systems have successfully contributed to a more detailed understanding of the isotropic-nematic transition in confinement \cite{Gruhn1998, Binder2008, Ji2009, Ji2009b}. A recent molecular dynamics study documented even radial movements of phase fronts analogous to the ones inferred here, however for a rod-like system in nanochannels\cite{Karjalainen2013}. Finally, diffraction experiments on partial fillings \cite{Huber2013} may allow for complementary structural information on the propagation of the phase boundaries in channel space and thus shed light on the remarkably different pathways found here for freezing and melting of discotics.

\section{Acknowledgement}
We are grateful to Julien Kelber and Harald Bock (CRPP) for having resynthesised and provided the liquid crystalline material studied here. This work has been supported by the Polish National Science Centre under the Project "Molecular Structure and Dynamics of Liquid Crystals Based Nanocomposites" (Decision No. DEC-2012/05/B/ST3/02782) and by the French-German Project TEMPLDISCO jointly founded by ANR and the German Science foundation (DFG SCHO 470/21-1 and HU 850/3-1).










\begin{thebibliography}{50}
\expandafter\ifx\csname natexlab\endcsname\relax\def\natexlab#1{#1}\fi
\expandafter\ifx\csname bibnamefont\endcsname\relax
  \def\bibnamefont#1{#1}\fi
\expandafter\ifx\csname bibfnamefont\endcsname\relax
  \def\bibfnamefont#1{#1}\fi
\expandafter\ifx\csname citenamefont\endcsname\relax
  \def\citenamefont#1{#1}\fi
\expandafter\ifx\csname url\endcsname\relax
  \def\url#1{\texttt{#1}}\fi
\expandafter\ifx\csname urlprefix\endcsname\relax\def\urlprefix{URL }\fi
\providecommand{\bibinfo}[2]{#2}
\providecommand{\eprint}[2][]{\url{#2}}

\bibitem[{\citenamefont{Laschat et~al.}(2007)\citenamefont{Laschat, Baro,
  Steinke, Giesselmann, Haegele, Scalia, Judele, Kapatsina, Sauer, Schreivogel
  et~al.}}]{Laschat2007}
\bibinfo{author}{\bibfnamefont{S.}~\bibnamefont{Laschat}},
  \bibinfo{author}{\bibfnamefont{A.}~\bibnamefont{Baro}},
  \bibinfo{author}{\bibfnamefont{N.}~\bibnamefont{Steinke}},
  \bibinfo{author}{\bibfnamefont{F.}~\bibnamefont{Giesselmann}},
  \bibinfo{author}{\bibfnamefont{C.}~\bibnamefont{Haegele}},
  \bibinfo{author}{\bibfnamefont{G.}~\bibnamefont{Scalia}},
  \bibinfo{author}{\bibfnamefont{R.}~\bibnamefont{Judele}},
  \bibinfo{author}{\bibfnamefont{E.}~\bibnamefont{Kapatsina}},
  \bibinfo{author}{\bibfnamefont{S.}~\bibnamefont{Sauer}},
  \bibinfo{author}{\bibfnamefont{A.}~\bibnamefont{Schreivogel}},
  \bibnamefont{et~al.}, \bibinfo{journal}{Angewandte Chemie-international
  Edition} \textbf{\bibinfo{volume}{46}}, \bibinfo{pages}{4832}
  (\bibinfo{year}{2007}).

\bibitem[{\citenamefont{Sergeyev et~al.}(2007)\citenamefont{Sergeyev, Pisula,
  and Geerts}}]{Sergeyev2007}
\bibinfo{author}{\bibfnamefont{S.}~\bibnamefont{Sergeyev}},
  \bibinfo{author}{\bibfnamefont{W.}~\bibnamefont{Pisula}}, \bibnamefont{and}
  \bibinfo{author}{\bibfnamefont{Y.~H.} \bibnamefont{Geerts}},
  \bibinfo{journal}{Chemical Society Reviews} \textbf{\bibinfo{volume}{36}},
  \bibinfo{pages}{1902} (\bibinfo{year}{2007}).

\bibitem[{\citenamefont{Bisoyi and Kumar}(2010)}]{Bisoyi2010}
\bibinfo{author}{\bibfnamefont{H.~K.} \bibnamefont{Bisoyi}} \bibnamefont{and}
  \bibinfo{author}{\bibfnamefont{S.}~\bibnamefont{Kumar}},
  \bibinfo{journal}{Chemical Society Reviews} \textbf{\bibinfo{volume}{39}},
  \bibinfo{pages}{264} (\bibinfo{year}{2010}).

\bibitem[{\citenamefont{Haverkate et~al.}(2011)\citenamefont{Haverkate, Zbiri,
  Johnson, Deme, Mulder, and Kearey}}]{Haverkate2011}
\bibinfo{author}{\bibfnamefont{L.~A.} \bibnamefont{Haverkate}},
  \bibinfo{author}{\bibfnamefont{M.}~\bibnamefont{Zbiri}},
  \bibinfo{author}{\bibfnamefont{M.~R.} \bibnamefont{Johnson}},
  \bibinfo{author}{\bibfnamefont{B.}~\bibnamefont{Deme}},
  \bibinfo{author}{\bibfnamefont{F.~M.} \bibnamefont{Mulder}},
  \bibnamefont{and} \bibinfo{author}{\bibfnamefont{G.~J.}
  \bibnamefont{Kearey}}, \bibinfo{journal}{Journal of Physical Chemistry B}
  \textbf{\bibinfo{volume}{115}}, \bibinfo{pages}{13809}
  (\bibinfo{year}{2011}).

\bibitem[{\citenamefont{Krause et~al.}(2012)\citenamefont{Krause, Yin,
  Cerclier, Morineau, Wurm, Schick, Emmerling, and Schoenhals}}]{Krause2012}
\bibinfo{author}{\bibfnamefont{C.}~\bibnamefont{Krause}},
  \bibinfo{author}{\bibfnamefont{H.}~\bibnamefont{Yin}},
  \bibinfo{author}{\bibfnamefont{C.}~\bibnamefont{Cerclier}},
  \bibinfo{author}{\bibfnamefont{D.}~\bibnamefont{Morineau}},
  \bibinfo{author}{\bibfnamefont{A.}~\bibnamefont{Wurm}},
  \bibinfo{author}{\bibfnamefont{C.}~\bibnamefont{Schick}},
  \bibinfo{author}{\bibfnamefont{F.}~\bibnamefont{Emmerling}},
  \bibnamefont{and}
  \bibinfo{author}{\bibfnamefont{A.}~\bibnamefont{Schoenhals}},
  \bibinfo{journal}{Soft Matter} \textbf{\bibinfo{volume}{8}},
  \bibinfo{pages}{11115} (\bibinfo{year}{2012}).

\bibitem[{\citenamefont{Schoenhals et~al.}(2014)\citenamefont{Schoenhals,
  Krause, Zorn, Emmerling, Frick, Huber, and Falkenhagen}}]{Schoenhals2014}
\bibinfo{author}{\bibfnamefont{A.}~\bibnamefont{Schoenhals}},
  \bibinfo{author}{\bibfnamefont{C.}~\bibnamefont{Krause}},
  \bibinfo{author}{\bibfnamefont{R.}~\bibnamefont{Zorn}},
  \bibinfo{author}{\bibfnamefont{F.}~\bibnamefont{Emmerling}},
  \bibinfo{author}{\bibfnamefont{B.}~\bibnamefont{Frick}},
  \bibinfo{author}{\bibfnamefont{P.}~\bibnamefont{Huber}}, \bibnamefont{and}
  \bibinfo{author}{\bibfnamefont{J.}~\bibnamefont{Falkenhagen}},
  \bibinfo{journal}{Phys. Chem. Chem. Phys}  (\bibinfo{year}{2014}).

\bibitem[{\citenamefont{Grelet and Bock}(2006)}]{Grelet2006}
\bibinfo{author}{\bibfnamefont{E.}~\bibnamefont{Grelet}} \bibnamefont{and}
  \bibinfo{author}{\bibfnamefont{H.}~\bibnamefont{Bock}},
  \bibinfo{journal}{Europhysics Letters} \textbf{\bibinfo{volume}{73}},
  \bibinfo{pages}{712} (\bibinfo{year}{2006}).

\bibitem[{\citenamefont{Feng et~al.}(2009)\citenamefont{Feng, Marcon, Pisula,
  Hansen, Kirkpatrick, Grozema, Andrienko, Kremer, and Muellen}}]{Feng2009}
\bibinfo{author}{\bibfnamefont{X.}~\bibnamefont{Feng}},
  \bibinfo{author}{\bibfnamefont{V.}~\bibnamefont{Marcon}},
  \bibinfo{author}{\bibfnamefont{W.}~\bibnamefont{Pisula}},
  \bibinfo{author}{\bibfnamefont{M.~R.} \bibnamefont{Hansen}},
  \bibinfo{author}{\bibfnamefont{J.}~\bibnamefont{Kirkpatrick}},
  \bibinfo{author}{\bibfnamefont{F.}~\bibnamefont{Grozema}},
  \bibinfo{author}{\bibfnamefont{D.}~\bibnamefont{Andrienko}},
  \bibinfo{author}{\bibfnamefont{K.}~\bibnamefont{Kremer}}, \bibnamefont{and}
  \bibinfo{author}{\bibfnamefont{K.}~\bibnamefont{Muellen}},
  \bibinfo{journal}{Nature Materials} \textbf{\bibinfo{volume}{8}},
  \bibinfo{pages}{421} (\bibinfo{year}{2009}).

\bibitem[{\citenamefont{Steinhart et~al.}(2005)\citenamefont{Steinhart,
  Zimmermann, Goring, Schaper, Gosele, Weder, and Wendorff}}]{Steinhart2005}
\bibinfo{author}{\bibfnamefont{M.}~\bibnamefont{Steinhart}},
  \bibinfo{author}{\bibfnamefont{S.}~\bibnamefont{Zimmermann}},
  \bibinfo{author}{\bibfnamefont{P.}~\bibnamefont{Goring}},
  \bibinfo{author}{\bibfnamefont{A.~K.} \bibnamefont{Schaper}},
  \bibinfo{author}{\bibfnamefont{U.}~\bibnamefont{Gosele}},
  \bibinfo{author}{\bibfnamefont{C.}~\bibnamefont{Weder}}, \bibnamefont{and}
  \bibinfo{author}{\bibfnamefont{J.~H.} \bibnamefont{Wendorff}},
  \bibinfo{journal}{Nano Letters} \textbf{\bibinfo{volume}{5}},
  \bibinfo{pages}{429} (\bibinfo{year}{2005}).

\bibitem[{\citenamefont{Duran et~al.}(2012)\citenamefont{Duran,
  Hartmann-Azanza, Steinhart, Gehrig, Laquai, Feng, Muellen, Butt, and
  Floudas}}]{Duran2012}
\bibinfo{author}{\bibfnamefont{H.}~\bibnamefont{Duran}},
  \bibinfo{author}{\bibfnamefont{B.}~\bibnamefont{Hartmann-Azanza}},
  \bibinfo{author}{\bibfnamefont{M.}~\bibnamefont{Steinhart}},
  \bibinfo{author}{\bibfnamefont{D.}~\bibnamefont{Gehrig}},
  \bibinfo{author}{\bibfnamefont{F.}~\bibnamefont{Laquai}},
  \bibinfo{author}{\bibfnamefont{X.}~\bibnamefont{Feng}},
  \bibinfo{author}{\bibfnamefont{K.}~\bibnamefont{Muellen}},
  \bibinfo{author}{\bibfnamefont{H.-J.} \bibnamefont{Butt}}, \bibnamefont{and}
  \bibinfo{author}{\bibfnamefont{G.}~\bibnamefont{Floudas}},
  \bibinfo{journal}{ACS Nano} \textbf{\bibinfo{volume}{6}},
  \bibinfo{pages}{9359} (\bibinfo{year}{2012}).

\bibitem[{\citenamefont{Crawford and Zumber}(1996)}]{Crawford1996}
\bibinfo{editor}{\bibfnamefont{G.}~\bibnamefont{Crawford}} \bibnamefont{and}
  \bibinfo{editor}{\bibfnamefont{S.}~\bibnamefont{Zumber}}, eds.,
  \emph{\bibinfo{title}{Liquid Crystals in Complex Geometries Formed by Polymer
  and Porous Networks}} (\bibinfo{publisher}{Taylor and Francis; New York,
  U.S.A.}, \bibinfo{year}{1996}).

\bibitem[{\citenamefont{Binder et~al.}(2008)\citenamefont{Binder, Horbach,
  Vink, and De~Virgiliis}}]{Binder2008}
\bibinfo{author}{\bibfnamefont{K.}~\bibnamefont{Binder}},
  \bibinfo{author}{\bibfnamefont{J.}~\bibnamefont{Horbach}},
  \bibinfo{author}{\bibfnamefont{R.}~\bibnamefont{Vink}}, \bibnamefont{and}
  \bibinfo{author}{\bibfnamefont{A.}~\bibnamefont{De~Virgiliis}},
  \bibinfo{journal}{Soft Matter} \textbf{\bibinfo{volume}{4}},
  \bibinfo{pages}{1555} (\bibinfo{year}{2008}).

\bibitem[{\citenamefont{Kityk et~al.}(2008)\citenamefont{Kityk, Wolff, Knorr,
  Morineau, Lefort, and Huber}}]{Kityk2008}
\bibinfo{author}{\bibfnamefont{A.~V.} \bibnamefont{Kityk}},
  \bibinfo{author}{\bibfnamefont{M.}~\bibnamefont{Wolff}},
  \bibinfo{author}{\bibfnamefont{K.}~\bibnamefont{Knorr}},
  \bibinfo{author}{\bibfnamefont{D.}~\bibnamefont{Morineau}},
  \bibinfo{author}{\bibfnamefont{R.}~\bibnamefont{Lefort}}, \bibnamefont{and}
  \bibinfo{author}{\bibfnamefont{P.}~\bibnamefont{Huber}},
  \bibinfo{journal}{Phys. Rev. Lett.} \textbf{\bibinfo{volume}{101}},
  \bibinfo{pages}{187801} (\bibinfo{year}{2008}).

\bibitem[{\citenamefont{Kopitzke et~al.}(2000)\citenamefont{Kopitzke, Wendorff,
  and Glusen}}]{Kopitzke2000}
\bibinfo{author}{\bibfnamefont{J.}~\bibnamefont{Kopitzke}},
  \bibinfo{author}{\bibfnamefont{J.~H.} \bibnamefont{Wendorff}},
  \bibnamefont{and} \bibinfo{author}{\bibfnamefont{B.}~\bibnamefont{Glusen}},
  \bibinfo{journal}{Liquid Crystals} \textbf{\bibinfo{volume}{27}},
  \bibinfo{pages}{643} (\bibinfo{year}{2000}).

\bibitem[{\citenamefont{Stillings et~al.}(2008)\citenamefont{Stillings, Martin,
  Steinhart, Pettau, Paraknowitsch, Geuss, Schmidt, Germano, Schmidt, Goesele
  et~al.}}]{Stillings2008}
\bibinfo{author}{\bibfnamefont{C.}~\bibnamefont{Stillings}},
  \bibinfo{author}{\bibfnamefont{E.}~\bibnamefont{Martin}},
  \bibinfo{author}{\bibfnamefont{M.}~\bibnamefont{Steinhart}},
  \bibinfo{author}{\bibfnamefont{R.}~\bibnamefont{Pettau}},
  \bibinfo{author}{\bibfnamefont{J.}~\bibnamefont{Paraknowitsch}},
  \bibinfo{author}{\bibfnamefont{M.}~\bibnamefont{Geuss}},
  \bibinfo{author}{\bibfnamefont{J.}~\bibnamefont{Schmidt}},
  \bibinfo{author}{\bibfnamefont{G.}~\bibnamefont{Germano}},
  \bibinfo{author}{\bibfnamefont{H.~.~W.} \bibnamefont{Schmidt}},
  \bibinfo{author}{\bibfnamefont{U.}~\bibnamefont{Goesele}},
  \bibnamefont{et~al.}, \bibinfo{journal}{Molecular Crystals and Liquid
  Crystals} \textbf{\bibinfo{volume}{495}}, \bibinfo{pages}{285}
  (\bibinfo{year}{2008}).

\bibitem[{\citenamefont{Hassheider et~al.}(2001)\citenamefont{Hassheider,
  Benning, Kitzerow, Achard, and Bock}}]{Hassheider2001}
\bibinfo{author}{\bibfnamefont{T.}~\bibnamefont{Hassheider}},
  \bibinfo{author}{\bibfnamefont{S.~A.} \bibnamefont{Benning}},
  \bibinfo{author}{\bibfnamefont{H.~S.} \bibnamefont{Kitzerow}},
  \bibinfo{author}{\bibfnamefont{M.~F.} \bibnamefont{Achard}},
  \bibnamefont{and} \bibinfo{author}{\bibfnamefont{H.}~\bibnamefont{Bock}},
  \bibinfo{journal}{Angewandte Chemie-international Edition}
  \textbf{\bibinfo{volume}{40}}, \bibinfo{pages}{2060} (\bibinfo{year}{2001}).

\bibitem[{\citenamefont{Gruener and Huber}(2008)}]{Gruener2008}
\bibinfo{author}{\bibfnamefont{S.}~\bibnamefont{Gruener}} \bibnamefont{and}
  \bibinfo{author}{\bibfnamefont{P.}~\bibnamefont{Huber}},
  \bibinfo{journal}{Phys. Rev. Lett.} \textbf{\bibinfo{volume}{100}},
  \bibinfo{pages}{064502} (\bibinfo{year}{2008}).

\bibitem[{\citenamefont{Huber et~al.}(2007)\citenamefont{Huber, Gruener,
  Schaefer, Knorr, and Kityk}}]{Huber2007}
\bibinfo{author}{\bibfnamefont{P.}~\bibnamefont{Huber}},
  \bibinfo{author}{\bibfnamefont{S.}~\bibnamefont{Gruener}},
  \bibinfo{author}{\bibfnamefont{C.}~\bibnamefont{Schaefer}},
  \bibinfo{author}{\bibfnamefont{K.}~\bibnamefont{Knorr}}, \bibnamefont{and}
  \bibinfo{author}{\bibfnamefont{A.~V.} \bibnamefont{Kityk}},
  \bibinfo{journal}{Eur. Phys. J. Special Topics}
  \textbf{\bibinfo{volume}{141}}, \bibinfo{pages}{101} (\bibinfo{year}{2007}).

\bibitem[{\citenamefont{Gruener et~al.}(2009)\citenamefont{Gruener, Hofmann,
  Wallacher, Kityk, and Huber}}]{Gruener2009}
\bibinfo{author}{\bibfnamefont{S.}~\bibnamefont{Gruener}},
  \bibinfo{author}{\bibfnamefont{T.}~\bibnamefont{Hofmann}},
  \bibinfo{author}{\bibfnamefont{D.}~\bibnamefont{Wallacher}},
  \bibinfo{author}{\bibfnamefont{A.~V.} \bibnamefont{Kityk}}, \bibnamefont{and}
  \bibinfo{author}{\bibfnamefont{P.}~\bibnamefont{Huber}},
  \bibinfo{journal}{Phys. Rev. E} \textbf{\bibinfo{volume}{79}},
  \bibinfo{pages}{067301} (\bibinfo{year}{2009}).

\bibitem[{\citenamefont{Gruener and Huber}(2011)}]{Gruener2011}
\bibinfo{author}{\bibfnamefont{S.}~\bibnamefont{Gruener}} \bibnamefont{and}
  \bibinfo{author}{\bibfnamefont{P.}~\bibnamefont{Huber}}, \bibinfo{journal}{J.
  Phys. : Cond. Matt.} \textbf{\bibinfo{volume}{23}}, \bibinfo{pages}{184109}
  (\bibinfo{year}{2011}).

\bibitem[{\citenamefont{Kityk et~al.}(2009)\citenamefont{Kityk, Knorr, and
  Huber}}]{Kityk2009}
\bibinfo{author}{\bibfnamefont{A.~V.} \bibnamefont{Kityk}},
  \bibinfo{author}{\bibfnamefont{K.}~\bibnamefont{Knorr}}, \bibnamefont{and}
  \bibinfo{author}{\bibfnamefont{P.}~\bibnamefont{Huber}},
  \bibinfo{journal}{Phys. Rev. B} \textbf{\bibinfo{volume}{80}},
  \bibinfo{pages}{035421} (\bibinfo{year}{2009}).

\bibitem[{\citenamefont{Brunet et~al.}(2011)\citenamefont{Brunet, Thiebaut,
  Charlet, Bock, Kelber, and Grelet}}]{Brunet2011}
\bibinfo{author}{\bibfnamefont{T.}~\bibnamefont{Brunet}},
  \bibinfo{author}{\bibfnamefont{O.}~\bibnamefont{Thiebaut}},
  \bibinfo{author}{\bibfnamefont{E.}~\bibnamefont{Charlet}},
  \bibinfo{author}{\bibfnamefont{H.}~\bibnamefont{Bock}},
  \bibinfo{author}{\bibfnamefont{J.}~\bibnamefont{Kelber}}, \bibnamefont{and}
  \bibinfo{author}{\bibfnamefont{E.}~\bibnamefont{Grelet}},
  \bibinfo{journal}{Epl} \textbf{\bibinfo{volume}{93}}, \bibinfo{pages}{16004}
  (\bibinfo{year}{2011}).

\bibitem[{\citenamefont{Cerclier et~al.}(2012)\citenamefont{Cerclier, Ndao,
  Busselez, Lefort, Grelet, Huber, Kityk, Noirez, Schoenhals, and
  Morineau}}]{Cerclier2012}
\bibinfo{author}{\bibfnamefont{C.~V.} \bibnamefont{Cerclier}},
  \bibinfo{author}{\bibfnamefont{M.}~\bibnamefont{Ndao}},
  \bibinfo{author}{\bibfnamefont{R.}~\bibnamefont{Busselez}},
  \bibinfo{author}{\bibfnamefont{R.}~\bibnamefont{Lefort}},
  \bibinfo{author}{\bibfnamefont{E.}~\bibnamefont{Grelet}},
  \bibinfo{author}{\bibfnamefont{P.}~\bibnamefont{Huber}},
  \bibinfo{author}{\bibfnamefont{A.~V.} \bibnamefont{Kityk}},
  \bibinfo{author}{\bibfnamefont{L.}~\bibnamefont{Noirez}},
  \bibinfo{author}{\bibfnamefont{A.}~\bibnamefont{Schoenhals}},
  \bibnamefont{and} \bibinfo{author}{\bibfnamefont{D.}~\bibnamefont{Morineau}},
  \bibinfo{journal}{Journal of Physical Chemistry C}
  \textbf{\bibinfo{volume}{116}}, \bibinfo{pages}{18990}
  (\bibinfo{year}{2012}).

\bibitem[{\citenamefont{Crawford et~al.}(1991)\citenamefont{Crawford, Allender,
  Doane, Vilfan, and Vilfan}}]{Crawford1991}
\bibinfo{author}{\bibfnamefont{G.~P.} \bibnamefont{Crawford}},
  \bibinfo{author}{\bibfnamefont{D.~W.} \bibnamefont{Allender}},
  \bibinfo{author}{\bibfnamefont{J.~W.} \bibnamefont{Doane}},
  \bibinfo{author}{\bibfnamefont{M.}~\bibnamefont{Vilfan}}, \bibnamefont{and}
  \bibinfo{author}{\bibfnamefont{I.}~\bibnamefont{Vilfan}},
  \bibinfo{journal}{Physical Review A} \textbf{\bibinfo{volume}{44}},
  \bibinfo{pages}{2570} (\bibinfo{year}{1991}).

\bibitem[{\citenamefont{Charlet and Grelet}(2008)}]{Charlet2008}
\bibinfo{author}{\bibfnamefont{E.}~\bibnamefont{Charlet}} \bibnamefont{and}
  \bibinfo{author}{\bibfnamefont{E.}~\bibnamefont{Grelet}},
  \bibinfo{journal}{Physical Review E} \textbf{\bibinfo{volume}{78}},
  \bibinfo{pages}{041707} (\bibinfo{year}{2008}).

\bibitem[{\citenamefont{Calus et~al.}(2012)\citenamefont{Calus, Rau, Huber, and
  Kityk}}]{Calus2012}
\bibinfo{author}{\bibfnamefont{S.}~\bibnamefont{Calus}},
  \bibinfo{author}{\bibfnamefont{D.}~\bibnamefont{Rau}},
  \bibinfo{author}{\bibfnamefont{P.}~\bibnamefont{Huber}}, \bibnamefont{and}
  \bibinfo{author}{\bibfnamefont{A.~V.} \bibnamefont{Kityk}},
  \bibinfo{journal}{Physical Review E} \textbf{\bibinfo{volume}{86}},
  \bibinfo{pages}{021701} (\bibinfo{year}{2012}).

\bibitem[{\citenamefont{Christenson}(2001)}]{Christenson2001}
\bibinfo{author}{\bibfnamefont{H.~K.} \bibnamefont{Christenson}},
  \bibinfo{journal}{Journal of Physics-condensed Matter}
  \textbf{\bibinfo{volume}{13}}, \bibinfo{pages}{R95} (\bibinfo{year}{2001}).

\bibitem[{\citenamefont{Wallacher et~al.}(2001)\citenamefont{Wallacher,
  Ackermann, Huber, Enderle, and Knorr}}]{Wallacher2001}
\bibinfo{author}{\bibfnamefont{D.}~\bibnamefont{Wallacher}},
  \bibinfo{author}{\bibfnamefont{R.}~\bibnamefont{Ackermann}},
  \bibinfo{author}{\bibfnamefont{P.}~\bibnamefont{Huber}},
  \bibinfo{author}{\bibfnamefont{M.}~\bibnamefont{Enderle}}, \bibnamefont{and}
  \bibinfo{author}{\bibfnamefont{K.}~\bibnamefont{Knorr}},
  \bibinfo{journal}{Phys. Rev. B} \textbf{\bibinfo{volume}{64}},
  \bibinfo{pages}{184203} (\bibinfo{year}{2001}).

\bibitem[{\citenamefont{Alba-Simionesco
  et~al.}(2006)\citenamefont{Alba-Simionesco, Coasne, Dosseh, Dudziak, Gubbins,
  Radhakrishnan, and Sliwinska-Bartkowiak}}]{Alba-Simionesco2006}
\bibinfo{author}{\bibfnamefont{C.}~\bibnamefont{Alba-Simionesco}},
  \bibinfo{author}{\bibfnamefont{B.}~\bibnamefont{Coasne}},
  \bibinfo{author}{\bibfnamefont{G.}~\bibnamefont{Dosseh}},
  \bibinfo{author}{\bibfnamefont{G.}~\bibnamefont{Dudziak}},
  \bibinfo{author}{\bibfnamefont{K.~E.} \bibnamefont{Gubbins}},
  \bibinfo{author}{\bibfnamefont{R.}~\bibnamefont{Radhakrishnan}},
  \bibnamefont{and}
  \bibinfo{author}{\bibfnamefont{M.}~\bibnamefont{Sliwinska-Bartkowiak}},
  \bibinfo{journal}{J. Phys.: Condens. Matter} \textbf{\bibinfo{volume}{18}},
  \bibinfo{pages}{R15} (\bibinfo{year}{2006}).

\bibitem[{\citenamefont{Knorr et~al.}(2008)\citenamefont{Knorr, Huber, and
  Wallacher}}]{Knorr2008}
\bibinfo{author}{\bibfnamefont{K.}~\bibnamefont{Knorr}},
  \bibinfo{author}{\bibfnamefont{P.}~\bibnamefont{Huber}}, \bibnamefont{and}
  \bibinfo{author}{\bibfnamefont{D.}~\bibnamefont{Wallacher}},
  \bibinfo{journal}{Zeitschrift für Physikalische Chemie}
  \textbf{\bibinfo{volume}{222}}, \bibinfo{pages}{257} (\bibinfo{year}{2008}).

\bibitem[{\citenamefont{Kityk and Huber}(2010)}]{Kityk2010}
\bibinfo{author}{\bibfnamefont{A.~V.} \bibnamefont{Kityk}} \bibnamefont{and}
  \bibinfo{author}{\bibfnamefont{P.}~\bibnamefont{Huber}},
  \bibinfo{journal}{Applied Physics Letters} \textbf{\bibinfo{volume}{97}},
  \bibinfo{pages}{153124} (\bibinfo{year}{2010}).

\bibitem[{\citenamefont{Krause and Schonhals}(2013)}]{schonhals2013}
\bibinfo{author}{\bibfnamefont{C.}~\bibnamefont{Krause}} \bibnamefont{and}
  \bibinfo{author}{\bibfnamefont{A.}~\bibnamefont{Schonhals}},
  \bibinfo{journal}{J. Phys. Chem. C} \textbf{\bibinfo{volume}{117}},
  \bibinfo{pages}{19712} (\bibinfo{year}{2013}).

\bibitem[{\citenamefont{Chahine et~al.}(2010)\citenamefont{Chahine, Kityk,
  Demarest, Jean, Knorr, Huber, Lefort, Zanotti, and Morineau}}]{Chahine2010}
\bibinfo{author}{\bibfnamefont{G.}~\bibnamefont{Chahine}},
  \bibinfo{author}{\bibfnamefont{A.~V.} \bibnamefont{Kityk}},
  \bibinfo{author}{\bibfnamefont{N.}~\bibnamefont{Demarest}},
  \bibinfo{author}{\bibfnamefont{F.}~\bibnamefont{Jean}},
  \bibinfo{author}{\bibfnamefont{K.}~\bibnamefont{Knorr}},
  \bibinfo{author}{\bibfnamefont{P.}~\bibnamefont{Huber}},
  \bibinfo{author}{\bibfnamefont{R.}~\bibnamefont{Lefort}},
  \bibinfo{author}{\bibfnamefont{J.-M.} \bibnamefont{Zanotti}},
  \bibnamefont{and} \bibinfo{author}{\bibfnamefont{D.}~\bibnamefont{Morineau}},
  \bibinfo{journal}{Physical review. E, Statistical, nonlinear, and soft matter
  physics} \textbf{\bibinfo{volume}{82}}, \bibinfo{pages}{011706}
  (\bibinfo{year}{2010}).

\bibitem[{\citenamefont{Oswald and Pieranski}(2006)}]{Oswald2006}
\bibinfo{author}{\bibfnamefont{P.}~\bibnamefont{Oswald}} \bibnamefont{and}
  \bibinfo{author}{\bibfnamefont{P.}~\bibnamefont{Pieranski}},
  \emph{\bibinfo{title}{Smectic and columnar liquid crystals}}
  (\bibinfo{publisher}{eds. G.W. Gray, J.W. Goodby, and A. Fukuda, CRC Press
  Taylor and Francis Group; Boca Raton, FL (USA)}, \bibinfo{year}{2006}).

\bibitem[{\citenamefont{Skacej et~al.}(1998)\citenamefont{Skacej,
  Alexe-Ionescu, Barbero, and Zumer}}]{Skacej1998}
\bibinfo{author}{\bibfnamefont{G.}~\bibnamefont{Skacej}},
  \bibinfo{author}{\bibfnamefont{A.~L.} \bibnamefont{Alexe-Ionescu}},
  \bibinfo{author}{\bibfnamefont{G.}~\bibnamefont{Barbero}}, \bibnamefont{and}
  \bibinfo{author}{\bibfnamefont{S.}~\bibnamefont{Zumer}},
  \bibinfo{journal}{Physical Review E} \textbf{\bibinfo{volume}{57}},
  \bibinfo{pages}{1780} (\bibinfo{year}{1998}).

\bibitem[{\citenamefont{Kutnjak et~al.}(2003)\citenamefont{Kutnjak, Kralj,
  Lahajnar, and Zumer}}]{Kutnjak2003}
\bibinfo{author}{\bibfnamefont{Z.}~\bibnamefont{Kutnjak}},
  \bibinfo{author}{\bibfnamefont{S.}~\bibnamefont{Kralj}},
  \bibinfo{author}{\bibfnamefont{G.}~\bibnamefont{Lahajnar}}, \bibnamefont{and}
  \bibinfo{author}{\bibfnamefont{S.}~\bibnamefont{Zumer}},
  \bibinfo{journal}{Phys. Rev. E} \textbf{\bibinfo{volume}{68}},
  \bibinfo{pages}{021705} (\bibinfo{year}{2003}).

\bibitem[{\citenamefont{Kutnjak et~al.}(2004)\citenamefont{Kutnjak, Kralj,
  Lahajnar, and Zumer}}]{Kutnjak2004}
\bibinfo{author}{\bibfnamefont{Z.}~\bibnamefont{Kutnjak}},
  \bibinfo{author}{\bibfnamefont{S.}~\bibnamefont{Kralj}},
  \bibinfo{author}{\bibfnamefont{G.}~\bibnamefont{Lahajnar}}, \bibnamefont{and}
  \bibinfo{author}{\bibfnamefont{S.}~\bibnamefont{Zumer}},
  \bibinfo{journal}{Phys. Rev. E} \textbf{\bibinfo{volume}{70}},
  \bibinfo{pages}{051703} (\bibinfo{year}{2004}).

\bibitem[{\citenamefont{Ocko et~al.}(1986)\citenamefont{Ocko, Braslau, Pershan,
  Alsnielsen, and Deutsch}}]{Ocko1986}
\bibinfo{author}{\bibfnamefont{B.~M.} \bibnamefont{Ocko}},
  \bibinfo{author}{\bibfnamefont{A.}~\bibnamefont{Braslau}},
  \bibinfo{author}{\bibfnamefont{P.~S.} \bibnamefont{Pershan}},
  \bibinfo{author}{\bibfnamefont{J.}~\bibnamefont{Alsnielsen}},
  \bibnamefont{and} \bibinfo{author}{\bibfnamefont{M.}~\bibnamefont{Deutsch}},
  \bibinfo{journal}{Phys. Rev. Lett.} \textbf{\bibinfo{volume}{57}},
  \bibinfo{pages}{94} (\bibinfo{year}{1986}), ISSN \bibinfo{issn}{{0031-9007}}.

\bibitem[{\citenamefont{Webber}(2010)}]{Webber2010}
\bibinfo{author}{\bibfnamefont{J.~B.~W.} \bibnamefont{Webber}},
  \bibinfo{journal}{Progress In Nuclear Magnetic Resonance Spectroscopy}
  \textbf{\bibinfo{volume}{56}}, \bibinfo{pages}{78} (\bibinfo{year}{2010}).

\bibitem[{\citenamefont{Schaefer et~al.}(2008)\citenamefont{Schaefer, Hofmann,
  Wallacher, Huber, and Knorr}}]{Schaefer2008}
\bibinfo{author}{\bibfnamefont{C.}~\bibnamefont{Schaefer}},
  \bibinfo{author}{\bibfnamefont{T.}~\bibnamefont{Hofmann}},
  \bibinfo{author}{\bibfnamefont{D.}~\bibnamefont{Wallacher}},
  \bibinfo{author}{\bibfnamefont{P.}~\bibnamefont{Huber}}, \bibnamefont{and}
  \bibinfo{author}{\bibfnamefont{K.}~\bibnamefont{Knorr}},
  \bibinfo{journal}{Phys. Rev. Lett.} \textbf{\bibinfo{volume}{100}},
  \bibinfo{pages}{175701} (\bibinfo{year}{2008}).

\bibitem[{\citenamefont{Moerz et~al.}(2012)\citenamefont{Moerz, Knorr, and
  Huber}}]{Moerz2012}
\bibinfo{author}{\bibfnamefont{S.~T.} \bibnamefont{Moerz}},
  \bibinfo{author}{\bibfnamefont{K.}~\bibnamefont{Knorr}}, \bibnamefont{and}
  \bibinfo{author}{\bibfnamefont{P.}~\bibnamefont{Huber}},
  \bibinfo{journal}{Phys. Rev. B} \textbf{\bibinfo{volume}{85}},
  \bibinfo{pages}{075403} (\bibinfo{year}{2012}).

\bibitem[{\citenamefont{Mitchell et~al.}(2008)\citenamefont{Mitchell, Webber,
  and Strange}}]{Mitchell2008}
\bibinfo{author}{\bibfnamefont{J.}~\bibnamefont{Mitchell}},
  \bibinfo{author}{\bibfnamefont{J.~B.~W.} \bibnamefont{Webber}},
  \bibnamefont{and} \bibinfo{author}{\bibfnamefont{J.~H.}
  \bibnamefont{Strange}}, \bibinfo{journal}{Physics Reports-review Section of
  Physics Letters} \textbf{\bibinfo{volume}{461}}, \bibinfo{pages}{1}
  (\bibinfo{year}{2008}).

\bibitem[{\citenamefont{Petrov and Furo}(2009)}]{Petrov2009}
\bibinfo{author}{\bibfnamefont{O.~V.} \bibnamefont{Petrov}} \bibnamefont{and}
  \bibinfo{author}{\bibfnamefont{I.}~\bibnamefont{Furo}},
  \bibinfo{journal}{Progress In Nuclear Magnetic Resonance Spectroscopy}
  \textbf{\bibinfo{volume}{54}}, \bibinfo{pages}{97} (\bibinfo{year}{2009}).

\bibitem[{\citenamefont{Riikonen et~al.}(2011)\citenamefont{Riikonen, Salonen,
  and Lehto}}]{Riikonen2011}
\bibinfo{author}{\bibfnamefont{J.}~\bibnamefont{Riikonen}},
  \bibinfo{author}{\bibfnamefont{J.}~\bibnamefont{Salonen}}, \bibnamefont{and}
  \bibinfo{author}{\bibfnamefont{V.~P.} \bibnamefont{Lehto}},
  \bibinfo{journal}{Journal of Thermal Analysis and Calorimetry}
  \textbf{\bibinfo{volume}{105}}, \bibinfo{pages}{823} (\bibinfo{year}{2011}).

\bibitem[{\citenamefont{Kondrashova and Valiullin}(2013)}]{Kondrashova2013}
\bibinfo{author}{\bibfnamefont{D.}~\bibnamefont{Kondrashova}} \bibnamefont{and}
  \bibinfo{author}{\bibfnamefont{R.}~\bibnamefont{Valiullin}},
  \bibinfo{journal}{Microporous and Mesoporous Materials}
  \textbf{\bibinfo{volume}{178}}, \bibinfo{pages}{15} (\bibinfo{year}{2013}).

\bibitem[{\citenamefont{Gruhn and Schoen}(1998)}]{Gruhn1998}
\bibinfo{author}{\bibfnamefont{T.}~\bibnamefont{Gruhn}} \bibnamefont{and}
  \bibinfo{author}{\bibfnamefont{M.}~\bibnamefont{Schoen}},
  \bibinfo{journal}{J. Chem. Phys.} \textbf{\bibinfo{volume}{108}},
  \bibinfo{pages}{9124} (\bibinfo{year}{1998}).

\bibitem[{\citenamefont{Ji et~al.}(2009{\natexlab{a}})\citenamefont{Ji, Lefort,
  Busselez, and Morineau}}]{Ji2009}
\bibinfo{author}{\bibfnamefont{Q.}~\bibnamefont{Ji}},
  \bibinfo{author}{\bibfnamefont{R.}~\bibnamefont{Lefort}},
  \bibinfo{author}{\bibfnamefont{R.}~\bibnamefont{Busselez}}, \bibnamefont{and}
  \bibinfo{author}{\bibfnamefont{D.}~\bibnamefont{Morineau}},
  \bibinfo{journal}{J. Chem. Phys.} \textbf{\bibinfo{volume}{130}},
  \bibinfo{pages}{234501} (\bibinfo{year}{2009}{\natexlab{a}}).

\bibitem[{\citenamefont{Ji et~al.}(2009{\natexlab{b}})\citenamefont{Ji, Lefort,
  and Morineau}}]{Ji2009b}
\bibinfo{author}{\bibfnamefont{Q.}~\bibnamefont{Ji}},
  \bibinfo{author}{\bibfnamefont{R.}~\bibnamefont{Lefort}}, \bibnamefont{and}
  \bibinfo{author}{\bibfnamefont{D.}~\bibnamefont{Morineau}},
  \bibinfo{journal}{Chem. Phys. Lett.} \textbf{\bibinfo{volume}{478}},
  \bibinfo{pages}{161} (\bibinfo{year}{2009}{\natexlab{b}}).

\bibitem[{\citenamefont{Karjalainen et~al.}(2013)\citenamefont{Karjalainen,
  Lintuvuori, Telkki, Lantto, and Vaara}}]{Karjalainen2013}
\bibinfo{author}{\bibfnamefont{J.}~\bibnamefont{Karjalainen}},
  \bibinfo{author}{\bibfnamefont{J.}~\bibnamefont{Lintuvuori}},
  \bibinfo{author}{\bibfnamefont{V.~V.} \bibnamefont{Telkki}},
  \bibinfo{author}{\bibfnamefont{P.}~\bibnamefont{Lantto}}, \bibnamefont{and}
  \bibinfo{author}{\bibfnamefont{J.}~\bibnamefont{Vaara}},
  \bibinfo{journal}{Physical Chemistry Chemical Physics}
  \textbf{\bibinfo{volume}{15}}, \bibinfo{pages}{14047} (\bibinfo{year}{2013}).

\bibitem[{\citenamefont{Huber et~al.}(2013)\citenamefont{Huber, Busch, Calus,
  and Kityk}}]{Huber2013}
\bibinfo{author}{\bibfnamefont{P.}~\bibnamefont{Huber}},
  \bibinfo{author}{\bibfnamefont{M.}~\bibnamefont{Busch}},
  \bibinfo{author}{\bibfnamefont{S.}~\bibnamefont{Calus}}, \bibnamefont{and}
  \bibinfo{author}{\bibfnamefont{A.~V.} \bibnamefont{Kityk}},
  \bibinfo{journal}{Physical Review E} \textbf{\bibinfo{volume}{87}},
  \bibinfo{pages}{042502} (\bibinfo{year}{2013}).

\end{thebibliography}

\end{document}